\documentclass[12pt]{article}
\usepackage{subcaption}
\usepackage{graphicx}
\usepackage{amsmath}
\usepackage[left=2cm, right=2cm, top=3cm]{geometry}

\usepackage[rightcaption]{sidecap}
\usepackage{subcaption}
\usepackage{graphicx}
\usepackage{amsfonts}
\usepackage{amsmath}
\usepackage{blkarray}
\usepackage{multirow}
\usepackage{arydshln}
\usepackage{graphicx}
\usepackage{subcaption}
\usepackage{float}
\usepackage{graphicx}
\usepackage{placeins}
\usepackage{float}

\newdimen\dummy
\dummy=\oddsidemargin
\begin{document}

\begin{titlepage}

\begin{center}
{\Large {\bf Darboux transformation and multi-soliton solutions of
discrete sine-Gordon equation}}\\\vspace{1in}

Y.~Hanif \footnote{Corresponding author: e-mail:
yasir\_pmc@yahoo.com } and U. Saleem \footnote{e-mail:
usaleem.physics@pu.edu.pk, usman\_physics@yahoo.com}

{{\it Department of Physics, University of the Punjab,\\
Quaid-e-Azam Campus, Lahore-54590, Pakistan.}}\\

\end{center}

\vspace{1cm}
\begin{abstract}
We study a discrete Darboux transformation and construct the
multi-soliton solutions in terms of ratio of determinants for
integrable discrete sine-Gordon equation. We also calculate explicit
expressions of single, double, triple, quad soliton solutions as
well as single and double breather solutions of discrete sine-Gordon
equation. Dynamical features of discrete kinks and breathers have
also been illustrated.
\end{abstract}

\vspace{1cm} PACS: 03.50.-z,02.30.Ik,
\end{titlepage}
\section{Introduction}

The integrable difference-difference, difference-differential
equations have attracted considerable attention in recent years. An
integrable discrete equation emerges as the superposition formula
for B\"{a}cklund transformations of its associated counterpart in
continuous regime. Most of the physical systems are described by
differential equations which reflect smoothness in the natural
processes and we face this situation during investigation of
macroscopic phenomena. On the other hand there exist a large number
of physical systems that are inherently discrete in nature and are
well described by difference equations rather than differential
equations. Mostly quantum mechanical systems are governed by
difference equations which under the continuum limit (if we take
step size infinitesimally small) could be reduced to associated
differential equations. Therefore, the discrete systems are more
elementary than their continuous counterparts. The discrete systems
have rich mathematical structure, actually the difference equations
are nonlocal before taking continuum limit. The nonlocality makes
such systems more complicated and difficult to explore integrability
of such systems. However, most of the established mathematical
techniques to explore interesting properties of continuous
integrable equations are no more applicable for the discrete
systems. For this reason it is necessary to develop classical
methods and discover new concepts to explore hidden mathematical
structures.

There has been a growing interest in the study of discrete
integrable systems. Some well known examples of discrete integrable
systems are Toda lattice system, nonlinear Schr\"{o}dinger equation
(also known as Ablowitz and Ladik system), Korteweg-deVries equation
(KdV), sine-Gordon, Volterra lattice system, etc. The sine-Gordon
(SG) equation has been widely used in different areas of physical
and mathematical sciences (for example see \cite{dsg01}-\cite{dsg02}
and references therein). First time SG equation was appeared in
theory of surfaces in differential geometry \cite{dsg12}. SG
equation describes the oscillations of coupled pendulum
\cite{dsg03}, propagation of magnetic-flux on Josephson array
\cite{dsg14}, dynamics of crystal dislocation \cite{dsg15}, also
explain the dynamics of DNA (deoxyribonucleic acid) double helix
molecule \cite{dsg17}. SG equation has attracted a great deal of
attention in recent decades.

SG equation is an integrable nonlinear equation and was solved by
the mean of inverse scattering transform (IST) method \cite{dsg05}.
The SG equation is defined as
\begin{equation}
\alpha_{xt}=\gamma\sin \alpha,\quad \quad \alpha=\alpha(x,t),
\label{sg1}
\end{equation}%
where $\gamma$ is a real constant and $\alpha$ be a continuous
scalar field. The partial derivative are denoted by subscripts. The
discrete integrable systems have multiple applications in different
fields from pure mathematics to applied sciences. The discrete
integrable equations also admit interesting properties such as
exactly solvable by IST method, existence of conservation laws,
admit multi-soliton solutions and so on \cite{dsr05}-\cite{dsr10}.

A semi-discrete sine-Gordon (sd-SG) equation is given by
\cite{dsr21}
\begin{equation}
\frac{d}{dt} \left(\alpha_{n+1}-\alpha_{n}\right)=4\gamma \sin
\frac{1}{2}\left( \alpha_{n+1}+\alpha_{n}\right).  \label{semisg5}
\end{equation}%

SG equation (both continuous and discrete) has a wide range
applications in different fields such as mathematics, physics and
life sciences. In particular SG equation has been widely used in the
study of nonlinear dynamics of DNA (deoxyribonucleic acid) double
helix molecule. In \cite{dsg17}, Yomosa has studied the dynamics DNA
double helix molecule under continuum limit and reduced the
associated problem into SG equation, however, the original governing
equation was discrete. So it is worthwhile to study the discrete SG
equation. In an earlier work \cite{Yasir2019}, we investigated
different solutions of a sd-SG equation (\ref{semisg5}). In this
paper, we would like to explore the dynamics of multi-soliton and
multi-breather solutions of discrete sine-Gordon (dSG) equation. The
solutions obtained in this article have not reported so far and each
expression reduces to the known solutions of semi-discrete and
continuous SG equations under continuum limits.

The article is organized as follow. Section 2 deals with a dSG
equation and associated discrete linear system. In section 3, we
apply Darboux transformation and obtain a determinant representation
of multi-soliton solutions. In Section 4, we obtain explicit
expressions of soliton solutions in zero background. We also plot
single, double, triple and quad soliton solutions as well as we
present the dynamics of single and double breather solutions. Last
section contains concluding remarks.

\section{Integrable discrete sine-Gordon equation}

An integrable dSG equation is given as
\begin{equation}
\sin\left(\frac{\alpha_{n+1,m+1}-\alpha_{n+1,m}-\alpha_{n,m+1}+\alpha_{n,m}}{4}\right)=\gamma
\sin
\left(\frac{\alpha_{n+1,m+1}+\alpha_{n+1,m}+\alpha_{n,m+1}+\alpha_{n,m}}{4}\right),
\label{sg5}
\end{equation}%
where $\alpha_{n,m}$ real discrete scalar field \cite{dsr18,dsr21}.
Under the continuum limits dSG equation (\ref{sg5}) reduces to sd-SG
equation (\ref{semisg5}) and classical SG equation (\ref{sg1}). In
order to obtain sd-SG equation (\ref{semisg5}) from (\ref{sg5}), we
define
\begin{equation}
\alpha_{i,j}=\alpha_{i}(t+\mathcal{T}j),\quad\quad
t=\mathcal{T}m,\quad\quad\mbox{and }
\gamma\rightarrow\mathcal{T}\gamma.\label{continuum1}
\end{equation}
Keeping $t$ fixed and applying the limit as $\mathcal{T}\rightarrow
0$ and $m \rightarrow \infty$, we get sd-SG equation
\begin{equation}
\frac{d}{dt} \left(\alpha_{n+1}-\alpha_{n}\right)=4\gamma \sin
\frac{1}{2}\left( \alpha_{n+1}+\alpha_{n}\right). \label{continuum2}
\end{equation}
Similarly, if we define
\begin{equation}
\alpha_{i}=\alpha(x+\mathcal{X}i),\quad\quad
x=\mathcal{X}n,\quad\quad\mbox{and}\quad
\gamma\rightarrow\mathcal{X}\gamma.\label{continuum3}
\end{equation}
If we keep $x$ fixed and apply the limit as $\mathcal{X}\rightarrow
0$ and $n \rightarrow \infty$, equation (\ref{continuum2}) reduces
to the classical SG equation (\ref{sg1})\footnote {If we define
\begin{equation}
\alpha_{i,j}=\alpha(x+\mathcal{X}i,t+\mathcal{T}j),\quad\quad
x=\mathcal{X}n,\quad t=\mathcal{T}m\quad\quad\mbox{with}\quad
\gamma\rightarrow\mathcal{XT}\gamma.\label{continuum}
\end{equation}
By keeping $x,t$ fixed and applying limit as $\mathcal{X}\rightarrow
0,\mathcal{T}\rightarrow 0$ and $n \rightarrow \infty$,$m
\rightarrow \infty$, equation (\ref{sg5}) yields the classical SG
equation (\ref{sg1})}.

Equation (\ref{sg5}) may be expressed as integrability condition of
following linear difference-difference equations for an auxiliary
function $\chi _{n,m}(\lambda)$ \cite{dsr21}
\begin{eqnarray}
{\cal N}\chi _{n,m}\equiv\chi _{n+1,m}&=&P_{n,m}\chi _{n,m}=(\lambda
T+U_{n,m})\chi _{n,m},
\label{sg6} \\
{\cal M}\chi _{n,m}\equiv\chi _{n,m+1} &=&Q_{n,m}\chi
_{n,m}=\left(I+\lambda ^{-1}V_{n,m}\right)\chi _{n,m},  \label{sg7}
\end{eqnarray}%
where ${\cal N}$ and ${\cal M}$ are shift operators in $n$ and $m$
respectively. The coefficient matrices are given by
\begin{eqnarray}
T &=&\left(
\begin{array}{cc}
0 & 1 \\
1 & 0%
\end{array}%
\right) ,\quad U_{n,m}=\left(
\begin{array}{cc}
e^{-\iota\left(\alpha_{n+1,m}-\alpha_{n,m}\right)/2}& 0 \\
0 &e^{\iota\left(\alpha_{n+1,m}-\alpha_{n,m}\right)/2}%
\end{array}%
\right) ,  \nonumber \\
I &=&\left(
\begin{array}{cc}
1 & 0 \\
0 & 1%
\end{array}%
\right) ,\quad V_{n,m} =\gamma \left(
\begin{array}{cc}
0 & e^{-\iota\left(\alpha_{n,m+1}+\alpha_{n,m}\right)/2} \\
e^{\iota\left(\alpha_{n,m+1}+\alpha_{n,m}\right)/2}& 0%
\end{array}%
\right),  \label{sg8}
\end{eqnarray}%
here $\chi _{n,m}=\left(
\begin{array}{cc}
X _{n,m}&
Y _{n,m}%
\end{array}%
\right)^{T}$ and $\iota=\sqrt{-1}$. The system of linear difference-difference equations (\ref{sg6})-(\ref{sg7}) is known as discrete linear system \cite%
{dsr18} and the associated integrability condition (${\cal M\,
N}\chi _{n,m}={\cal N\,M}\chi _{n,m}$) gives a discrete zero
curvature condition (that is, $P_{n,m+1}Q_{n,m}=Q_{n+1,m}P_{n,m}$).
Again under continuum limits linear system (\ref{sg6})-(\ref{sg7})
reduces into linear eigenvalue problem for sd-SG equation
(\ref{semisg5}) and continuous SG equation (\ref{sg1}).
\section{Darboux transformation}
The Darboux transformation is the most powerful method to generate
an infinite series of solutions of linear equations (differential as
well as difference equations) from a known solution \cite{dsr24}.
The one-fold Darboux transformation for the system
(\ref{sg6})-(\ref{sg7}) can be defined as
\begin{eqnarray}
X _{n,m}[1] &=&\lambda Y _{n,m}-\Gamma^{(1)} _{n,m}X_{n,m},  \label{sd-DT1a} \\
Y_{n,m}[1] &=&\lambda X_{n,m}-\Omega^{(1)} _{n,m}Y_{n,m},
\label{sd-DT2a}
\end{eqnarray}
the unknown coefficients $\Gamma^{(1)} _{n,m}$ and $\Omega^{(1)}
_{n,m}$ may be obtained from the following conditions
\begin{eqnarray}
\left. X _{n,m}[1]\right|_{\lambda=\lambda_{1}, X _{n,m}=X^{(1)}_{n,m},Y _{n,m}=Y^{(1)}_{n,m}}&=&0,  \label{sd-DT1b} \\
\left. Y _{n,m}[1]\right|_{\lambda=\lambda_{1}, X
_{n,m}=X^{(1)}_{n,m},Y _{n,m}=Y^{(1)}_{n,m}}&=&0. \label{sd-DT2b}
\end{eqnarray}
The above conditions allow us to write one-fold transformation
(\ref{sd-DT1a})-(\ref{sd-DT2a}) as
\begin{eqnarray}
X _{n,m}[1] &\equiv&\lambda Y _{n,m}-\frac{\lambda _{1}Y^{(1)}
_{n,m}}{X^{(1)}_{n,m}}%
X_{n,m}=\frac{\det\left(
                \begin{array}{cc}
 \lambda Y _{n,m} & X_{n,m} \\
 \lambda _{1}Y^{(1)}
_{n,m}& X^{(1)}
_{n,m} \\
                \end{array}
              \right)
}{X^{(1)}_{n,m}},  \label{sd-DT1} \\
Y_{n,m}[1] &\equiv&\lambda X_{n,m}-\frac{\lambda _{1}X^{(1)}_{n,m}}{Y^{(1)}_{n,m}}%
Y_{n,m}=\frac{\det\left(
                \begin{array}{cc}
 \lambda X _{n,m} & Y_{n,m} \\
 \lambda _{1}X^{(1)}
_{n,m}& Y^{(1)}
_{n,m} \\
                \end{array}
              \right)
}{Y^{(1)}_{n,m}},  \label{sd-DT2}
\end{eqnarray}
here $X^{(1)}_{n,m}$ and $Y^{(1)}_{n,m}$ denote the particular
solutions to the linear system (\ref{sg6})-(\ref{sg7}). The linear
system of difference-difference equations (\ref{sg6})-(\ref{sg7}) is
covariant under the action of Darboux transformation
(\ref{sd-DT1})-(\ref{sd-DT2}), that is,
\begin{eqnarray}
X_{n+1,m}[1] &=&e^{-\iota\left(\alpha_{n+1,m}[1]-\alpha_{n,m}[1]\right)/2}X_{n,m}[1]+\lambda Y_{n,m}[1],  \nonumber \\
Y_{n+1,m}[1] &=&\lambda X_{n,m}[1]+e^{\iota\left(\alpha_{n+1,m}[1]-\alpha_{n,m}[1]\right)/2}Y_{n,m}[1],  \label{sd-Lax1 DT} \\
X _{n,m+1}[1] &=& X_{n,m}[1]+\lambda ^{-1}\gamma e^{-\iota\left(\alpha_{n,m+1}[1]+\alpha_{n,m}[1]\right)/2}Y_{n,m}[1],\qquad \qquad  \nonumber \\
Y_{n,m+1} [1] &=&\lambda ^{-1}\gamma
e^{\iota\left(\alpha_{n,m+1}[1]+\alpha_{n,m}[1]\right)/2}X_{n,m}[1]+Y_{n,m}[1].
\label{sd-Lax2 DT}
\end{eqnarray}%
By substitution of scalar functions $X_{n,m}[1]$ and $Y_{n,m}[1]$
from (\ref{sd-DT1})-(\ref{sd-DT2}) in (\ref{sd-Lax1
DT})-(\ref{sd-Lax2 DT}), we obtain
\begin{equation}
e^{-\iota\left(\alpha_{n+1,m}[1]-\alpha_{n,m}[1]\right)/2}=
e^{-\iota\left(\alpha_{n+1,m}-\alpha_{n,m}\right)/2}\left( \frac{Y _{n+1,m}^{(1)}}{X _{n+1,m}^{(1)}}\right)\left( \frac{X _{n,m}^{(1)}}{Y _{n,m}^{(1)}}%
\right), \label{sg15}
\end{equation}%
which implies,%
\begin{equation}
\alpha_{n,m}\left[ 1\right] =\alpha_{n,m}+2\iota\ln \left( \frac{Y
_{n,m}^{(1)}}{X _{n,m}^{(1)}}\right). \label{sg16}
\end{equation}%
The two-fold Darboux transformation is defined as
\begin{eqnarray}
X _{n,m}[2] &\equiv&\lambda Y _{n,m}[1]-\Gamma^{(2)}_{n,m}[1]X_{n,m}[1]=\lambda^2X_{n,m}-f^{1}_{n,m}\lambda Y_{n,m}-f^{0}_{n,m}X_{n,m},  \label{Twosd-DT1a} \\
Y_{n,m}[2] &\equiv&\lambda X_{n,m}[1]-\Omega^{(2)}
_{n,m}[1]Y_{n,m}[1]=\lambda^2Y_{n,m}-g^{1}_{n,m}\lambda
X_{n,m}-g^{0}_{n,m}Y_{n,m}, \label{Twosd-DT2a}
\end{eqnarray}
with $f^{1}_{n,m}=\Omega^{(1)}_{n,m}+\Gamma^{(2)}_{n,m}[1],
f^{0}_{n,m}=-\Gamma^{(2)}_{n,m}[1]\Gamma^{(1)}_{n,m}$ and
$g^{1}_{n,m}=\Gamma^{(1)}_{n,m}+\Omega^{(2)}_{n,m}[1],
g^{0}_{n,m}=-\Omega^{(2)}_{n,m}[1]\Omega^{(1)}_{n,m}$. The unknown
coefficients $f^{0}_{n,m},f^{1}_{n,m}$ and $g^{0}_{n,m},g^{1}_{n,m}$
can be determined from the following conditions
\begin{eqnarray}
\left. X _{n,m}[2]\right|_{\lambda=\lambda_{k}, X _{n,m}=X^{(k)}_{n,m},Y _{n,m}=Y^{(k)}_{n,m}}&=&0,  \label{Twosd-DT1b} \\
\left. Y _{n,m}[2]\right|_{\lambda=\lambda_{k}, X
_{n,m}=X^{(k)}_{n,m},Y _{n,m}=Y^{(k)}_{n,m}}&=&0, \label{Twosd-DT2b}
\end{eqnarray}
for $k=1,2$. The above conditions reduce to following systems of
linear equations
\begin{eqnarray}
\left(
  \begin{array}{cc}
    X^{(1)}_{n,m} & \lambda_{1}Y^{(1)}_{n,m} \\
    X^{(2)}_{n,m} & \lambda_{2}Y^{(2)}_{n,m} \\
  \end{array}
\right)\left(
         \begin{array}{c}
           f^{0}_{n,m} \\
           f^{1}_{n,m} \\
         \end{array}
       \right)
&=&\left(
      \begin{array}{c}
        \lambda_{1}^2X^{(1)}_{n,m} \\
         \lambda_{2}^2X^{(2)}_{n,m} \\
      \end{array}
    \right)
,  \label{Twosd-DT1c} \\
\left(
  \begin{array}{cc}
    Y^{(1)}_{n,m} & \lambda_{1}X^{(1)}_{n,m} \\
    Y^{(2)}_{n,m} & \lambda_{2}X^{(2)}_{n,m} \\
  \end{array}
\right)\left(
         \begin{array}{c}
           g^{0}_{n,m} \\
           g^{1}_{n,m} \\
         \end{array}
       \right)
&=&\left(
      \begin{array}{c}
        \lambda_{1}^2Y^{(1)}_{n,m} \\
         \lambda_{2}^2Y^{(2)}_{n,m} \\
      \end{array}
    \right)
.\label{Twosd-DT2c}
\end{eqnarray}
On substitution of unknown coefficients the two-fold transformation
(\ref{Twosd-DT1a})-(\ref{Twosd-DT2a}) reduces to
\begin{eqnarray}
X _{n,m}[2] &=&\frac{\det\left(
                \begin{array}{ccc}
 \lambda^{2}X _{n,m} &\lambda Y _{n,m} &X_{n,m}\\
 \lambda_{1}^{2}X^{(1)} _{n,m} &\lambda_{1} Y^{(1)} _{n,m} &X^{(1)}_{n,m} \\
 \lambda_{2}^{2}X^{(2)} _{n,m} &\lambda_{2} Y^{(2)} _{n,m} &X^{(2)}_{n,m}
                \end{array}
              \right)
}{\det\left(
                \begin{array}{cc}
  \lambda_{1} Y^{(1)} _{n,m} &X^{(1)}_{n,m} \\
 \lambda_{2} Y^{(2)} _{n,m} &X^{(2)}_{n,m}
                \end{array}
              \right)},  \label{Twosd-DT1} \\
Y_{n,m}[2] &=&\frac{\det\left(
                \begin{array}{ccc}
 \lambda^{2}Y _{n,m} &\lambda X _{n,m} &Y_{n,m}\\
 \lambda_{1}^{2}Y^{(1)} _{n,m} &\lambda_{1} X^{(1)} _{n,m} &Y^{(1)}_{n,m} \\
 \lambda_{2}^{2}Y^{(2)} _{n,m} &\lambda_{2} X^{(2)} _{n,m} &Y^{(2)}_{n,m}
                \end{array}
              \right)
}{\det\left(
                \begin{array}{cc}
  \lambda_{1} X^{(1)} _{n,m} &Y^{(1)}_{n,m} \\
 \lambda_{2} X^{(2)} _{n,m} &Y^{(2)}_{n,m}
                \end{array}
              \right)}.  \label{Twosd-DT2}
\end{eqnarray}
The two-fold transformed solutions given by
(\ref{Twosd-DT1})-(\ref{Twosd-DT2}) also satisfy the linear system
of difference-difference equations (\ref{sg6})-(\ref{sg7})
\begin{eqnarray}
X_{n+1,m}[2] &=&e^{-\iota\left(\alpha_{n+1,m}[2]-\alpha_{n,m}[2]\right)/2}X_{n,m}[2]+\lambda Y_{n,m}[2],  \nonumber \\
Y_{n+1,m}[2] &=&\lambda X_{n,m}[2]+e^{\iota\left(\alpha_{n+1,m}[2]-\alpha_{n,m}[2]\right)/2}Y_{n,m}[2],  \label{1Twosd-Lax1 DT} \\
X _{n,m+1}[2] &=& X_{n,m}[2]+\lambda ^{-1}\gamma e^{-\iota\left(\alpha_{n,m+1}[2]+\alpha_{n,m}[2]\right)/2}Y_{n,m}[2],  \nonumber \\
Y_{n,m+1} [2] &=&\lambda ^{-1}\gamma
e^{\iota\left(\alpha_{n,m+1}[2]+\alpha_{n,m}[2]\right)/2}X_{n,m}[2]+Y_{n,m}[2].
\label{1Twosd-Lax2 DT}
\end{eqnarray}%
Using equations (\ref{Twosd-DT1})-(\ref{Twosd-DT2}) in equations
(\ref{1Twosd-Lax1 DT})-(\ref{1Twosd-Lax2 DT}), we obtain
\begin{equation}
e^{-\iota\left(\alpha_{n+1,m}[2]-\alpha_{n,m}[2]\right)/2}=
e^{-\iota\left(\alpha_{n+1,m}-\alpha_{n,m}\right)/2}\left(
\frac{f^{0}_{n+1,m}}{f^{0}_{n,m}}\right), \label{Twosg15}
\end{equation}%
which yields,%
\begin{equation}
\alpha_{n,m}\left[2\right]=\alpha_{n,m}+2\iota\ln\frac{\det\left(
                \begin{array}{cc}
  \lambda_{1} X^{(1)} _{n,m} &Y^{(1)}_{n,m} \\
 \lambda_{2} X^{(2)} _{n,m} &Y^{(2)}_{n,m}
                \end{array}
              \right)}{\det\left(
                \begin{array}{cc}
  \lambda_{1}Y^{(1)}_{n,m} &X^{(1)} _{n,m}\\
\lambda_{2}Y^{(2)}_{n,m}&X^{(2)} _{n,m}
                \end{array}
              \right)}. \label{Twosg16}
\end{equation}
Similarly, three-fold Darboux transformation is defined as
\begin{eqnarray}
X _{n,m}[3] &\equiv&\lambda Y _{n,m}[2]-\Gamma^{(3)}_{n,m}[2]X_{n,m}[2]=\lambda^3Y_{n,m}-f^{2}_{n,m}\lambda^2 X_{n,m}-f^{1}_{n,m}\lambda Y_{n,m}-f^{0}_{n,m} X_{n,m},  \label{Threesd-DT1a} \\
Y_{n,m}[2] &\equiv&\lambda X
_{n,m}[2]-\Omega^{(3)}_{n,m}[2]Y_{n,m}[2]=\lambda^3X_{n,m}-g^{2}_{n,m}\lambda^2
Y_{n,m}-g^{1}_{n,m}\lambda X_{n,m}-g^{0}_{n,m} Y_{n,m}.
\label{Threesd-DT2a}
\end{eqnarray}
The unknown coefficients $f^{0}_{n,m},f^{1}_{n,m},f^{2}_{n,m}$ and
$g^{0}_{n,m},g^{1}_{n,m},g^{2}_{n,m}$ can be determined from the
following conditions
\begin{eqnarray}
\left. X _{n,m}[3]\right|_{\lambda=\lambda_{k}, X _{n,m}=X^{(k)}_{n,m},Y _{n,m}=Y^{(k)}_{n,m}}&=&0,  \label{Threesd-DT1b} \\
\left. Y _{n,m}[3]\right|_{\lambda=\lambda_{k}, X
_{n,m}=X^{(k)}_{n,m},Y _{n,m}=Y^{(k)}_{n,m}}&=&0,
\label{Threesd-DT2b}
\end{eqnarray}
for $k=1,2,3$. The above conditions can also be written as
\begin{eqnarray}
\left(
  \begin{array}{ccc}
    X^{(1)}_{n,m} & \lambda_{1}Y^{(1)}_{n,m} &\lambda^{2}_{1}X^{(1)}_{n,m}\\
    X^{(2)}_{n,m} & \lambda_{2}Y^{(2)}_{n,m} &\lambda^{2}_{2}X^{(2)}_{n,m}\\
     X^{(3)}_{n,m} & \lambda_{3}Y^{(3)}_{n,m} &\lambda^{2}_{3}X^{(3)}_{n,m}

  \end{array}
\right)\left(
         \begin{array}{c}
           f^{0}_{n,m} \\
           f^{1}_{n,m} \\
           f^{2}_{n,m}
         \end{array}
       \right)
&=&\left(
      \begin{array}{c}
        \lambda_{1}^3Y^{(1)}_{n,m} \\
         \lambda_{2}^3Y^{(2)}_{n,m} \\
         \lambda_{3}^3Y^{(3)}_{n,m}
      \end{array}
    \right)
,  \label{Threesd-DT1c} \\
\left(
  \begin{array}{ccc}
    Y^{(1)}_{n,m} & \lambda_{1}X^{(1)}_{n,m} &\lambda^{2}_{1}Y^{(1)}_{n,m}\\
    Y^{(2)}_{n,m} & \lambda_{2}X^{(2)}_{n,m} &\lambda^{2}_{2}Y^{(2)}_{n,m}\\
     Y^{(3)}_{n,m} & \lambda_{3}X^{(3)}_{n,m} &\lambda^{2}_{3}Y^{(3)}_{n,m}

  \end{array}
\right)\left(
         \begin{array}{c}
           g^{0}_{n,m} \\
           g^{1}_{n,m} \\
           g^{2}_{n,m}
         \end{array}
       \right)
&=&\left(
      \begin{array}{c}
        \lambda_{1}^3X^{(1)}_{n,m} \\
         \lambda_{2}^3X^{(2)}_{n,m} \\
         \lambda_{3}^3X^{(3)}_{n,m}
      \end{array}
    \right)
 .\label{Threesd-DT2c}
\end{eqnarray}
After substitution of unknown coefficients the expression of
three-fold transformation (\ref{Threesd-DT1a})-(\ref{Threesd-DT2a})
can also be expressed as ratio of determinants
\begin{eqnarray}
X _{n,m}[3] &=&\frac{\det\left(
                \begin{array}{cccc}
 \lambda^{3}Y _{n,m}&\lambda^{2} X _{n,m} &\lambda Y _{n,m} &X_{n,m}\\
 \lambda_{1}^{3}Y^{(1)} _{n,m}&\lambda_{1}^{2} X^{(1)} _{n,m} &\lambda_{1} Y^{(1)} _{n,m} &X^{(1)}_{n,m}\\
 \lambda_{2}^{3}Y^{(2)} _{n,m}&\lambda_{2}^{2} X^{(2)} _{n,m} &\lambda_{2} Y^{(2)}_{n,m}&X^{(2)}_{n,m}\\
 \lambda_{3}^{3}Y^{(3)} _{n,m}&\lambda_{3}^{2} X^{(3)} _{n,m} &\lambda_{3} Y^{(3)} _{n,m} &X^{(3)}_{n,m}
                \end{array}
              \right)
}{\det\left(
                \begin{array}{ccc}
   \lambda_{1}^{2} X^{(1)} _{n,m} &\lambda_{1} Y^{(1)} _{n,m} &X^{(1)}_{n,m}\\
 \lambda_{2}^{2} X^{(2)} _{n,m} &\lambda_{2} Y^{(2)}_{n,m}&X^{(2)}_{n,m}\\
 \lambda_{3}^{2} X^{(3)} _{n,m} &\lambda_{3} Y^{(3)} _{n,m} &X^{(3)}_{n,m}
                \end{array}
              \right)},  \label{Threesd-DT1} \\
Y_{n,m}[3] &=&\frac{\det\left(
                \begin{array}{cccc}
  \lambda^{3}X _{n,m}&\lambda^{2} Y _{n,m} &\lambda X _{n,m} &Y_{n,m}\\
 \lambda_{1}^{3}X^{(1)} _{n,m}&\lambda_{1}^{2} Y^{(1)} _{n,m} &\lambda_{1} X^{(1)} _{n,m} &Y^{(1)}_{n,m}\\
 \lambda_{2}^{3}X^{(2)} _{n,m}&\lambda_{2}^{2} Y^{(2)} _{n,m} &\lambda_{2} X^{(2)}_{n,m}&Y^{(2)}_{n,m}\\
 \lambda_{3}^{3}X^{(3)} _{n,m}&\lambda_{3}^{2} Y^{(3)} _{n,m} &\lambda_{3} X^{(3)} _{n,m} &Y^{(3)}_{n,m}
                \end{array}
              \right)
}{\det\left(
                \begin{array}{ccc}
  \lambda_{1}^{2} Y^{(1)} _{n,m} &\lambda_{1} X^{(1)} _{n,m} &Y^{(1)}_{n,m}\\
 \lambda_{2}^{2} Y^{(2)} _{n,m} &\lambda_{2} X^{(2)}_{n,m}&Y^{(2)}_{n,m}\\
 \lambda_{3}^{2} Y^{(3)} _{n,m} &\lambda_{3} X^{(3)} _{n,m} &Y^{(3)}_{n,m}
                \end{array}
              \right)}.  \label{Threesd-DT2}
\end{eqnarray}
Three-fold transformed solutions $X _{n,m}[3]$ and $Y _{n,m}[3]$
satisfy the linear system of difference-difference equations
(\ref{sg6})-(\ref{sg7}) such as
\begin{eqnarray}
X_{n+1,m}[3] &=&e^{-\iota\left(\alpha_{n+1,m}[3]-\alpha_{n,m}[3]\right)/2}X_{n,m}[3]+\lambda Y_{n,m}[3],  \nonumber \\
Y_{n+1,m}[3] &=&\lambda X_{n,m}[3]+e^{\iota\left(\alpha_{n+1,m}[3]-\alpha_{n,m}[3]\right)/2}Y_{n,m}[3],  \label{Threesd-Lax1 DT} \\
X _{n,m+1}[3] &=& X_{n,m}[3]+\lambda ^{-1}\gamma e^{-\iota\left(\alpha_{n,m+1}[3]+\alpha_{n,m}[3]\right)/2}Y_{n,m}[3],  \nonumber \\
Y_{n,m+1} [3] &=&\lambda ^{-1}\gamma
e^{\iota\left(\alpha_{n,m+1}[3]+\alpha_{n,m}[3]\right)/2}X_{n,m}[3]+Y_{n,m}[3].
\label{Threesd-Lax2 DT}
\end{eqnarray}%
Using equations (\ref{Threesd-DT1})-(\ref{Threesd-DT2}) in
(\ref{Threesd-Lax1 DT})-(\ref{Threesd-Lax2 DT}), we obtain
\begin{equation}
e^{-\iota\left(\alpha_{n+1,m}[3]-\alpha_{n,m}[3]\right)/2}=
e^{-\iota\left(\alpha_{n+1,m}-\alpha_{n,m}\right)/2}\left(
\frac{f^{0}_{n+1,m}}{f^{0}_{n,m}}\right), \label{Threesg15}
\end{equation}%
above equation gives%
\begin{equation}
\alpha_{n,m}\left[3\right]=\alpha_{n,m}+2\iota\ln\frac{\det\left(
                \begin{array}{ccc}
  \lambda_{1}^{2} Y^{(1)} _{n,m} &\lambda_{1} X^{(1)} _{n,m} &Y^{(1)}_{n,m} \\
 \lambda_{2}^{2} Y^{(2)} _{n,m} & \lambda_{2} X^{(2)}
 _{n,m}&Y^{(2)}_{n,m}\\
 \lambda_{3}^{2} Y^{(3)} _{n,m} & \lambda_{3} X^{(3)} _{n,m}&Y^{(3)}_{n,m}
                \end{array}
              \right)}{\det\left(
                \begin{array}{ccc}
    \lambda_{1}^{2} X^{(1)} _{n,m} &\lambda_{1} Y^{(1)} _{n,m} &X^{(1)}_{n,m} \\
 \lambda_{2}^{2} X^{(2)} _{n,m} & \lambda_{2} Y^{(2)} _{n,m}&X^{(2)}_{n,m}\\
 \lambda_{3}^{2} X^{(3)} _{n,m} & \lambda_{3} Y^{(3)} _{n,m}&X^{(3)}_{n,m}
                \end{array}
              \right)}. \label{Threesg16}
\end{equation}%

In what follows, we would like to generalize our results for
$N$-times iteration of Darboux transformation. For this purpose,
first take $N=2K$, the Darboux transformation can be factorize as
\begin{eqnarray}
X _{n,m}[2K] &=&\lambda^{2K}X_{n,m}-f^{2K-1}_{n,m}\lambda^{2K-1} Y_{n,m}-\dots -f^{1}_{n,m}\lambda Y_{n,m}-f^{0}_{n,m}X_{n,m},  \label{TwoKsd-DT1a} \\
Y_{n,m}[2K]
&=&\lambda^{2K}Y_{n,m}-g^{2K-1}_{n,m}\lambda^{2K-1}X_{n,m}-\dots-g^{1}_{n,m}\lambda
X_{n,m}-g^{0}_{n,m}Y_{n,m},\label{TwoKsd-DT2a}
\end{eqnarray}
where the unknown coefficients can be computed from the following
conditions
\begin{eqnarray}
\left. X _{n,m}[2K]\right|_{\lambda=\lambda_{k}, X _{n,m}=X^{(k)}_{n,m},Y _{n,m}=Y^{(k)}_{n,m}}&=&0,  \label{TwoKsd-DT1b} \\
\left. Y _{n,m}[2K]\right|_{\lambda=\lambda_{k}, X
_{n,m}=X^{(k)}_{n,m},Y _{n,m}=Y^{(k)}_{n,m}}&=&0,
\label{TwoKsd-DT2b}
\end{eqnarray}
for $k=1,2,\dots, 2K$. The above conditions reduce to following
systems of linear equations \scriptsize
\begin{eqnarray}
\left(
  \begin{array}{ccccc}
    X^{(1)}_{n,m} & \lambda_{1}Y^{(1)}_{n,m}&\dots &\lambda_{1}^{2K-2}X^{(1)}_{n,m} & \lambda_{1}^{2K-1}Y^{(1)}_{n,m} \\
    X^{(2)}_{n,m} & \lambda_{2} Y^{(2)}_{n,m}&\dots &\lambda_{2}^{2K-2}X^{(2)}_{n,m} & \lambda_{2}^{2K-1}Y^{(2)}_{n,m} \\
    \vdots&\vdots&\ddots&\vdots&\vdots\\
    X^{(2K-1)}_{n,m} &  \lambda_{2K-1}Y^{(2K-1)}_{n,m}&\dots &\lambda_{2K-1}^{2K-2}X^{(2K-1)}_{n,m} & \lambda_{2K-1}^{2K-1}Y^{(2K-1)}_{n,m} \\
    X^{(2K)}_{n,m} &  \lambda_{2K}Y^{(2K)}_{n,m}&\dots &\lambda_{2K}^{2K-2}X^{(2K)}_{n,m} & \lambda_{2K}^{2K-1}Y^{(2K)}_{n,m}
    \end{array}
\right)\left(
         \begin{array}{c}
           f^{0}_{n,m} \\
           f^{1}_{n,m} \\
           \vdots\\
           f^{2K-2}_{n,m} \\
           f^{2K-1}_{n,m} \\

         \end{array}
       \right)
&=&\left(
      \begin{array}{c}
        \lambda_{1}^{2K}X^{(1)}_{n,m} \\
         \lambda_{2}^{2K}X^{(2)}_{n,m} \\
           \vdots\\
           \lambda_{2K-1}^{2K}X^{(2K-1)}_{n,m} \\
         \lambda_{2K}^{2K}X^{(2K)}_{n,m} \\
      \end{array}
    \right)
, \notag\\ \label{TwoKsd-DT1c} \\
\left(
  \begin{array}{ccccc}
    Y^{(1)}_{n,m} & \lambda_{1}X^{(1)}_{n,m}&\dots &\lambda_{1}^{2K-2}Y^{(1)}_{n,m} & \lambda_{1}^{2K-1}X^{(1)}_{n,m} \\
    Y^{(2)}_{n,m} & \lambda_{2}X^{(2)}_{n,m}&\dots &\lambda_{2}^{2K-2}Y^{(2)}_{n,m} & \lambda_{2}^{2K-1}X^{(2)}_{n,m} \\
    \vdots&\vdots&\ddots&\vdots&\vdots\\
    Y^{(2K-1)}_{n,m} & \lambda_{2K-1}X^{(2K-1)}_{n,m}&\dots &\lambda_{2K-1}^{2K-2}Y^{(2K-1)}_{n,m} & \lambda_{2K-1}^{2K-1}X^{(2K-1)}_{n,m} \\
    Y^{(2K)}_{n,m} & \lambda_{2K}X^{(2K)}_{n,m}&\dots &\lambda_{2K}^{2K-2}Y^{(2K)}_{n,m} & \lambda_{2K}^{2K-1}X^{(2K)}_{n,m}
    \end{array}
\right)\left(
         \begin{array}{c}
           g^{0}_{n,m} \\
           g^{1}_{n,m} \\
           \vdots\\
           g^{2K-2}_{n,m} \\
           g^{2K-1}_{n,m} \\

         \end{array}
       \right)
&=&\left(
      \begin{array}{c}
        \lambda_{1}^{2K}Y^{(1)}_{n,m} \\
         \lambda_{2}^{2K}Y^{(2)}_{n,m} \\
           \vdots\\
           \lambda_{2K-1}^{2K}Y^{(2K-1)}_{n,m} \\
         \lambda_{2K}^{2K}Y^{(2K)}_{n,m} \\
      \end{array}
    \right)
, \notag\\.\label{TwoKsd-DT2c}
\end{eqnarray}
\normalsize
 Using the values of unknown coefficients the $2K$-fold transformation becomes (\ref{TwoKsd-DT1a})-(\ref{TwoKsd-DT2a}) as
\begin{eqnarray}
X _{n,m}[2K] &=&\frac{\det\left(
                \begin{array}{ccccc}
 \lambda^{2K}X _{n,m} &\lambda^{2K-1} Y _{n,m}&\dots& \lambda Y_{n,m}&X_{n,m}\\
 \lambda_{1}^{2K}X^{(1)} _{n,m} &\lambda_{1}^{2K-1} Y^{(1)}_{n,m}&\dots& \lambda_{1}Y^{(1)}_{n,m}&X^{(1)}_{n,m}\\
 \vdots&\vdots&\ddots&\vdots&\vdots\\
 \lambda_{2K-1}^{2K}X^{(2K-1)} _{n,m} &\lambda_{2K-1}^{2K-1} Y^{(2K-1)}_{n,m}&\dots&
 \lambda_{2K-1}Y^{(2K-1)}_{n,m}&X^{(2K-1)}_{n,m}\\
 \lambda_{2K}^{2K}X^{(2K)} _{n,m} &\lambda_{2K}^{2K-1} Y^{(2K)}_{n,m}&\dots& \lambda_{2K}Y^{(2K)}_{n,m}&X^{(2K)}_{n,m}
                \end{array}
              \right)
}{\det\left(
                \begin{array}{cccccccc}
 \lambda_{1}^{2K-1} Y^{(1)}_{n,m}& \lambda_{1}^{2K-2} X^{(1)}_{n,m}&\dots& \lambda_{1}Y^{(1)}_{n,m}&X^{(1)}_{n,m}\\
 \lambda_{2}^{2K-1} Y^{(2)}_{n,m}&\lambda_{2}^{2K-2} X^{(2)}_{n,m}&\dots& \lambda_{2}Y^{(2)}_{n,m}&X^{(2)}_{n,m}\\
 \vdots&\vdots&\ddots&\vdots&\vdots\\
 \lambda_{2K-1}^{2K-1} Y^{(2K-1)}_{n,m}&\lambda_{2K-1}^{2K-2} X^{(2K-1)}_{n,m}&\dots&
 \lambda_{2K-1}Y^{(2K-1)}_{n,m}&X^{(2K-1)}_{n,m}\\
 \lambda_{2K}^{2K-1} Y^{(2K)}_{n,m}&\lambda_{2K}^{2K-2} X^{(2K)}_{n,m}&\dots& \lambda_{2K}Y^{(2K)}_{n,m}&X^{(2K)}_{n,m}
                \end{array}
              \right)},  \label{TwoKsd-DT1} \\
Y_{n,m}[2K] &=&\frac{\det\left(
                \begin{array}{ccccc}
 \lambda^{2K}Y _{n,m} &\lambda^{2K-1} X _{n,m}&\dots& \lambda X_{n,m}&Y_{n,m}\\
 \lambda_{1}^{2K}Y^{(1)} _{n,m} &\lambda_{1}^{2K-1} X^{(1)}_{n,m}&\dots& \lambda_{1}X^{(1)}_{n,m}&Y^{(1)}_{n,m}\\
 \vdots&\vdots&\ddots&\vdots&\vdots\\
 \lambda_{2K-1}^{2K}Y^{(2K-1)} _{n,m} &\lambda_{2K-1}^{2K-1} X^{(2K-1)}_{n,m}&\dots&
 \lambda_{2K-1}X^{(2K-1)}_{n,m}&Y^{(2K-1)}_{n,m}\\
 \lambda_{2K}^{2K}Y^{(2K)} _{n,m} &\lambda_{2K}^{2K-1} X^{(2K)}_{n,m}&\dots& \lambda_{2K}X^{(2K)}_{n,m}&Y^{(2K)}_{n,m}
                \end{array}
              \right)
}{\det\left(
                \begin{array}{cccccccc}
 \lambda_{1}^{2K-1} X^{(1)}_{n,m}& \lambda_{1}^{2K-2} Y^{(1)}_{n,m}&\dots& \lambda_{1}X^{(1)}_{n,m}&Y^{(1)}_{n,m}\\
 \lambda_{2}^{2K-1} X^{(2)}_{n,m}&\lambda_{2}^{2K-2} Y^{(2)}_{n,m}&\dots& \lambda_{2}X^{(2)}_{n,m}&Y^{(2)}_{n,m}\\
 \vdots&\vdots&\ddots&\vdots&\vdots\\
 \lambda_{2K-1}^{2K-1} X^{(2K-1)}_{n,m}&\lambda_{2K-1}^{2K-2} Y^{(2K-1)}_{n,m}&\dots&
 \lambda_{2K-1}X^{(2K-1)}_{n,m}&Y^{(2K-1)}_{n,m}\\
 \lambda_{2K}^{2K-1} X^{(2K)}_{n,m}&\lambda_{2K}^{2K-2} Y^{(2K)}_{n,m}&\dots& \lambda_{2K}X^{(2K)}_{n,m}&Y^{(2K)}_{n,m}
                \end{array}
              \right)}.  \label{TwoKsd-DT2}
\end{eqnarray}
$2K$-fold transformed dynamical variable is given by
\begin{eqnarray}
\alpha_{n,m}\left[2K\right]
&=&\alpha_{n,m}+2\iota\ln\frac{\det\left(
                \begin{array}{cccccccc}
 \lambda_{1}^{2K-1} X^{(1)}_{n,m}& \lambda_{1}^{2K-2} Y^{(1)}_{n,m}&\dots& \lambda_{1}X^{(1)}_{n,m}&Y^{(1)}_{n,m}\\
 \lambda_{2}^{2K-1} X^{(2)}_{n,m}&\lambda_{2}^{2K-2} Y^{(2)}_{n,m}&\dots& \lambda_{2}X^{(2)}_{n,m}&Y^{(2)}_{n,m}\\
 \vdots&\vdots&\ddots&\vdots&\vdots\\
 \lambda_{2K-1}^{2K-1} X^{(2K-1)}_{n,m}&\lambda_{2K-1}^{2K-2} Y^{(2K-1)}_{n,m}&\dots&
 \lambda_{2K-1}X^{(2K-1)}_{n,m}&Y^{(2K-1)}_{n,m}\\
 \lambda_{2K}^{2K-1} X^{(2K)}_{n,m}&\lambda_{2K}^{2K-2} Y^{(2K)}_{n,m}&\dots& \lambda_{2K}X^{(2K)}_{n,m}&Y^{(2K)}_{n,m}
                \end{array}
              \right)}{\det\left(
                \begin{array}{cccccccc}
 \lambda_{1}^{2K-1} Y^{(1)}_{n,m}& \lambda_{1}^{2K-2} X^{(1)}_{n,m}&\dots& \lambda_{1}Y^{(1)}_{n,m}&X^{(1)}_{n,m}\\
 \lambda_{2}^{2K-1} Y^{(2)}_{n,m}&\lambda_{2}^{2K-2} X^{(2)}_{n,m}&\dots& \lambda_{2}Y^{(2)}_{n,m}&X^{(2)}_{n,m}\\
 \vdots&\vdots&\ddots&\vdots&\vdots\\
 \lambda_{2K-1}^{2K-1} Y^{(2K-1)}_{n,m}&\lambda_{2K-1}^{2K-2} X^{(2K-1)}_{n,m}&\dots&
 \lambda_{2K-1}Y^{(2K-1)}_{n,m}&X^{(2K-1)}_{n,m}\\
 \lambda_{2K}^{2K-1} Y^{(2K)}_{n,m}&\lambda_{2K}^{2K-2} X^{(2K)}_{n,m}&\dots& \lambda_{2K}Y^{(2K)}_{n,m}&X^{(2K)}_{n,m}
                \end{array}
              \right)}.\notag\\ \label{2Kfold-sg16}
\end{eqnarray}%

Similarly, for $N=2K+1$, we have
\begin{eqnarray}
X _{n,m}[2K+1] &=&\lambda^{2K+1}Y_{n,m}-f^{2K}_{n,m}\lambda^{2K} X_{n,m}-\dots -f^{1}_{n,m}\lambda Y_{n,m}-f^{0}_{n,m}X_{n,m},  \label{TwoKsd-DT1a} \\
Y_{n,m}[2K+1]
&=&\lambda^{2K+1}X_{n,m}-g^{2K}_{n,m}\lambda^{2K}Y_{n,m}-\dots-g^{1}_{n,m}\lambda
X_{n,m}-g^{0}_{n,m}Y_{n,m},\label{TwoKasd-DT2a}
\end{eqnarray}
along with following conditions
\begin{eqnarray}
\left. X _{n,m}[2K+1]\right|_{\lambda=\lambda_{k}, X _{n,m}=X^{(k)}_{n,m},Y _{n,m}=Y^{(k)}_{n,m}}&=&0,  \label{TwoKsd-DT1b} \\
\left. Y _{n,m}[2K+1]\right|_{\lambda=\lambda_{k}, X
_{n,m}=X^{(k)}_{n,m},Y _{n,m}=Y^{(k)}_{n,m}}&=&0,
\label{TwoKsd-DT2b}
\end{eqnarray}
for $k=1,2,\dots,2K+1$. The above conditions can also be expressed
in matrix notation as  \scriptsize
\begin{eqnarray}
\left(
  \begin{array}{ccccc}
    X^{(1)}_{n,m} & \lambda_{1}Y^{(1)}_{n,m}&\dots &\lambda_{1}^{2K-1}Y^{(1)}_{n,m} & \lambda_{1}^{2K}X^{(1)}_{n,m} \\
    X^{(2)}_{n,m} & \lambda_{2}Y^{(2)}_{n,m}&\dots &\lambda_{2}^{2K-1}Y^{(2)}_{n,m} & \lambda_{2}^{2K}X^{(2)}_{n,m} \\
    \vdots&\vdots&\ddots&\vdots&\vdots\\
    X^{(2K)}_{n,m} &\lambda_{2K} Y^{(2K)}_{n,m}&\dots &\lambda_{2K}^{2K-1}Y^{(2K-1)}_{n,m} & \lambda_{2K}^{2K}X^{(2K)}_{n,m} \\
    X^{(2K+1)}_{n,m} & \lambda_{2K+1}Y^{(2K+1)}_{n,m}&\dots &\lambda_{2K+1}^{2K-1}Y^{(2K+1)}_{n,m} & \lambda_{2K+1}^{2K}X^{(2K+1)}_{n,m}
    \end{array}
\right)\left(
         \begin{array}{c}
           f^{0}_{n,m} \\
           f^{1}_{n,m} \\
           \vdots\\
           f^{2K-1}_{n,m} \\
           f^{2K}_{n,m} \\

         \end{array}
       \right)
&=&\left(
      \begin{array}{c}
        \lambda_{1}^{2K+1}Y^{(1)}_{n,m} \\
         \lambda_{2}^{2K+1}Y^{(2)}_{n,m} \\
           \vdots\\
           \lambda_{2K}^{2K+1}Y^{(2K)}_{n,m} \\
         \lambda_{2K+1}^{2K+1}Y^{(2K+1)}_{n,m} \\
      \end{array}
    \right)
, \notag\\ \label{TwoKsd-DT1c} \\
\left(
  \begin{array}{ccccc}
    Y^{(1)}_{n,m} & \lambda_{1}X^{(1)}_{n,m}&\dots &\lambda_{1}^{2K-1}X^{(1)}_{n,m} & \lambda_{1}^{2K}Y^{(1)}_{n,m} \\
    Y^{(2)}_{n,m} & \lambda_{2}X^{(2)}_{n,m}&\dots &\lambda_{2}^{2K-1}X^{(2)}_{n,m} & \lambda_{2}^{2K}Y^{(2)}_{n,m} \\
    \vdots&\vdots&\ddots&\vdots&\vdots\\
    Y^{(2K)}_{n,m} &\lambda_{2K} X^{(2K)}_{n,m}&\dots &\lambda_{2K}^{2K-1}X^{(2K)}_{n,m} & \lambda_{2K}^{2K}Y^{(2K)}_{n,m} \\
    Y^{(2K+1)}_{n,m} & \lambda_{2K+1}X^{(2K+1)}_{n,m}&\dots &\lambda_{2K+1}^{2K-1}X^{(2K+1)}_{n,m} & \lambda_{2K+1}^{2K}Y^{(2K+1)}_{n,m}
    \end{array}
\right)\left(
         \begin{array}{c}
           g^{0}_{n,m} \\
           g^{1}_{n,m} \\
           \vdots\\
           g^{2K-1}_{n,m} \\
           g^{2K}_{n,m} \\

         \end{array}
       \right)
&=&\left(
      \begin{array}{c}
        \lambda_{1}^{2K+1}X^{(1)}_{n,m} \\
         \lambda_{2}^{2K+1}X^{(2)}_{n,m} \\
           \vdots\\
           \lambda_{2K}^{2K+1}X^{(2K)}_{n,m} \\
         \lambda_{2K+1}^{2K+1}X^{(2K+1)}_{n,m} \\
      \end{array}
    \right)
, \notag\\.\label{TwoKsd-DT2c}
\end{eqnarray}
\normalsize Using the values of unknown coefficients the
$(2K+1)$-fold transformation (\ref{TwoKsd-DT1a})-(\ref{TwoKsd-DT2a})
can be expressed as
\begin{eqnarray}
X _{n,m}[2K+1] &=&\frac{\det\left(
                \begin{array}{ccccc}
 \lambda^{2K+1}Y _{n,m} &\lambda^{2K} X _{n,m}&\dots& \lambda Y_{n,m}&X_{n,m}\\
 \lambda_{1}^{2K+1}Y^{(1)} _{n,m} &\lambda_{1}^{2K} X^{(1)}_{n,m}&\dots& \lambda_{1}Y^{(1)}_{n,m}&X^{(1)}_{n,m}\\
 \vdots&\vdots&\ddots&\vdots&\vdots\\
 \lambda_{2K}^{2K+1}Y^{(2K)} _{n,m} &\lambda_{2K}^{2K} X^{(2K)}_{n,m}&\dots&
 \lambda_{2K}Y^{(2K)}_{n,m}&X^{(2K)}_{n,m}\\
 \lambda_{2K+1}^{2K+1}Y^{(2K+1)} _{n,m} &\lambda_{2K+1}^{2K} X^{(2K+1)}_{n,m}&\dots& \lambda_{2K+1}Y^{(2K+1)}_{n,m}&X^{(2K+1)}_{n,m}
                \end{array}
              \right)
}{\det\left(
                \begin{array}{ccccc}
\lambda_{1}^{2K} X^{(1)}_{n,m}&\lambda_{1}^{2K-1} Y^{(1)}_{n,m}&\dots& \lambda_{1}Y^{(1)}_{n,m}&X^{(1)}_{n,m}\\
 \lambda_{2}^{2K} X^{(2)}_{n,m}&\lambda_{2}^{2K-1} Y^{(2)}_{n,m}&\dots& \lambda_{2}Y^{(2)}_{n,m}&X^{(2)}_{n,m}\\
 \vdots&\vdots&\ddots&\vdots&\vdots\\
 \lambda_{2K}^{2K} X^{(2K)}_{n,m}& \lambda_{2K}^{2K-1} Y^{(2K)}_{n,m}&\dots&
 \lambda_{2K}Y^{(2K)}_{n,m}&X^{(2K)}_{n,m}\\
 \lambda_{2K+1}^{2K} X^{(2K+1)}_{n,m}&\lambda_{2K+1}^{2K-1} Y^{(2K+1)}_{n,m}&\dots& \lambda_{2K+1}Y^{(2K+1)}_{n,m}&X^{(2K+1)}_{n,m}
                \end{array}
              \right)},  \label{TwoK+1sd-DT1} \\
Y_{n,m}[2K+1] &=&\frac{\det\left(
                \begin{array}{ccccc}
 \lambda^{2K+1}X _{n,m} &\lambda^{2K} Y _{n,m}&\dots& \lambda X_{n,m}&Y_{n,m}\\
 \lambda_{1}^{2K+1}X^{(1)} _{n,m} &\lambda_{1}^{2K} Y^{(1)}_{n,m}&\dots& \lambda_{1}X^{(1)}_{n,m}&Y^{(1)}_{n,m}\\
 \vdots&\vdots&\ddots&\vdots&\vdots\\
 \lambda_{2K}^{2K+1}X^{(2K)} _{n,m} &\lambda_{2K}^{2K} Y^{(2K)}_{n,m}&\dots&
 \lambda_{2K}X^{(2K)}_{n,m}&Y^{(2K)}_{n,m}\\
 \lambda_{2K+1}^{2K+1}X^{(2K+1)} _{n,m} &\lambda_{2K+1}^{2K} Y^{(2K+1)}_{n,m}&\dots& \lambda_{2K+1}X^{(2K+1)}_{n,m}&Y^{(2K+1)}_{n,m}
                \end{array}
              \right)
}{\det\left(
                \begin{array}{ccccc}
\lambda_{1}^{2K} Y^{(1)}_{n,m}&\lambda_{1}^{2K-1} X^{(1)}_{n,m}&\dots& \lambda_{1}X^{(1)}_{n,m}&Y^{(1)}_{n,m}\\
 \lambda_{2}^{2K} Y^{(2)}_{n,m}&\lambda_{2}^{2K-1} X^{(2)}_{n,m}&\dots& \lambda_{2}X^{(2)}_{n,m}&Y^{(2)}_{n,m}\\
 \vdots&\vdots&\ddots&\vdots&\vdots\\
 \lambda_{2K}^{2K} X^{(2K)}_{n,m}& \lambda_{2K}^{2K-1} X^{(2K)}_{n,m}&\dots&
 \lambda_{2K}X^{(2K)}_{n,m}&Y^{(2K)}_{n,m}\\
 \lambda_{2K+1}^{2K} X^{(2K+1)}_{n,m}&\lambda_{2K+1}^{2K-1} X^{(2K+1)}_{n,m}&\dots& \lambda_{2K+1}X^{(2K+1)}_{n,m}&Y^{(2K+1)}_{n,m}
                \end{array}
              \right)}.  \label{TwoK+1sd-DT2}
\end{eqnarray}
The $(2K+1)$-fold transformed solutions given by
(\ref{TwoK+1sd-DT1})-(\ref{TwoK+1sd-DT2}) satisfy the linear system
of difference-difference equations (\ref{sg6})-(\ref{sg7})
\begin{eqnarray}
X_{n+1,m}[2K+1] &=&e^{-\iota\left(\alpha_{n+1,m}[2K+1]-\alpha_{n,m}[2K+1]\right)/2}X_{n,m}[2K+1]+\lambda Y_{n,m}[2K+1],  \nonumber \\
Y_{n+1,m}[2K+1] &=&\lambda X_{n,m}[2K+1]+e^{\iota\left(\alpha_{n+1,m}[2K+1]-\alpha_{n,m}[2K+1]\right)/2}Y_{n,m}[2K+1],  \label{TwosdK+1-Lax1 DT} \\
X _{n,m+1}[2K+1] &=& X_{n,m}[2K+1]+\lambda ^{-1}\gamma e^{-\iota\left(\alpha_{n,m+1}[2K+1]+\alpha_{n,m}[2K+1]\right)/2}Y_{n,m}[2K+1],  \nonumber \\
Y_{n,m+1} [2K+1] &=&\lambda ^{-1}\gamma
e^{\iota\left(\alpha_{n,m+1}[2K+1]+\alpha_{n,m}[2K+1]\right)/2}X_{n,m}[2K+1]+Y_{n,m}[2K+1].
\label{TwosdK+1-Lax2 DT}
\end{eqnarray}%
Using equations (\ref{TwoK+1sd-DT1})-(\ref{TwoK+1sd-DT2}) in
(\ref{TwosdK+1-Lax1 DT})-(\ref{TwosdK+1-Lax2 DT}), we obtain
\begin{equation}
\alpha_{n,m}\left[2K+1\right]=\alpha_{n,m}+2\iota\ln\frac{\det\left(
                \begin{array}{cccccc}
 \lambda_{1}^{2K}Y^{(1)} _{n,m} &\lambda_{1}^{2K-1} X^{(1)}_{n,m}&\dots& \lambda_{1}X^{(1)}_{n,m}&Y^{(1)}_{n,m}\\
 \lambda_{2}^{2K}Y^{(2)} _{n,m} &\lambda_{2}^{2K-1} X^{(2)}_{n,m}&\dots& \lambda_{2}X^{(2)}_{n,m}&Y^{(2)}_{n,m}\\
 \vdots&\vdots&\ddots&\vdots&\vdots\\
 \lambda_{2K}^{2K}Y^{(2K)} _{n,m} &\lambda_{2K}^{2K-1} X^{(2K)}_{n,m}&\dots&
 \lambda_{2K}X^{(2K)}_{n,m}&Y^{(2K)}_{n,m}\\
 \lambda_{2K+1}^{2K}Y^{(2K+1)} _{n,m} &\lambda_{2K+1}^{2K-1} X^{(2K+1)}_{n,m}&\dots& \lambda_{2K+1}X^{(2K+1)}_{n,m}&Y^{(2K+1)}_{n,m}
                \end{array}
              \right)}{\det\left(
                \begin{array}{cccccc}
 \lambda_{1}^{2K}X^{(1)} _{n,m} &\lambda_{1}^{2K-1} Y^{(1)}_{n,m}&\dots& \lambda_{1}Y^{(1)}_{n,m}&X^{(1)}_{n,m}\\
 \lambda_{2}^{2K}X^{(2)} _{n,m} &\lambda_{2}^{2K-1} Y^{(2)}_{n,m}&\dots& \lambda_{2}Y^{(2)}_{n,m}&X^{(2)}_{n,m}\\
 \vdots&\vdots&\ddots&\vdots&\vdots\\
 \lambda_{2K}^{2K}X^{(2K)} _{n,m} &\lambda_{2K}^{2K-1} Y^{(2K)}_{n,m}&\dots&
 \lambda_{2K}Y^{(2K)}_{n,m}&X^{(2K)}_{n,m}\\
 \lambda_{2K+1}^{2K}X^{(2K+1)} _{n,m} &\lambda_{2K+1}^{2K-1} Y^{(2K+1)}_{n,m}&\dots& \lambda_{2K+1}Y^{(2K+1)}_{n,m}&X^{(2K+1)}_{n,m}
                \end{array}
              \right)}. \label{2K+1fold-sg16}
\end{equation}%
In the following section, we shall derive explicit expressions of
single, double, triple and quad soliton solutions and illustrate our
results for different choices of parameters.
\section{Explicit solutions and their dynamics}

\begin{figure}[H]
   \centering
    \begin{subfigure}[b]{0.45\textwidth}
       \includegraphics[width=\textwidth]{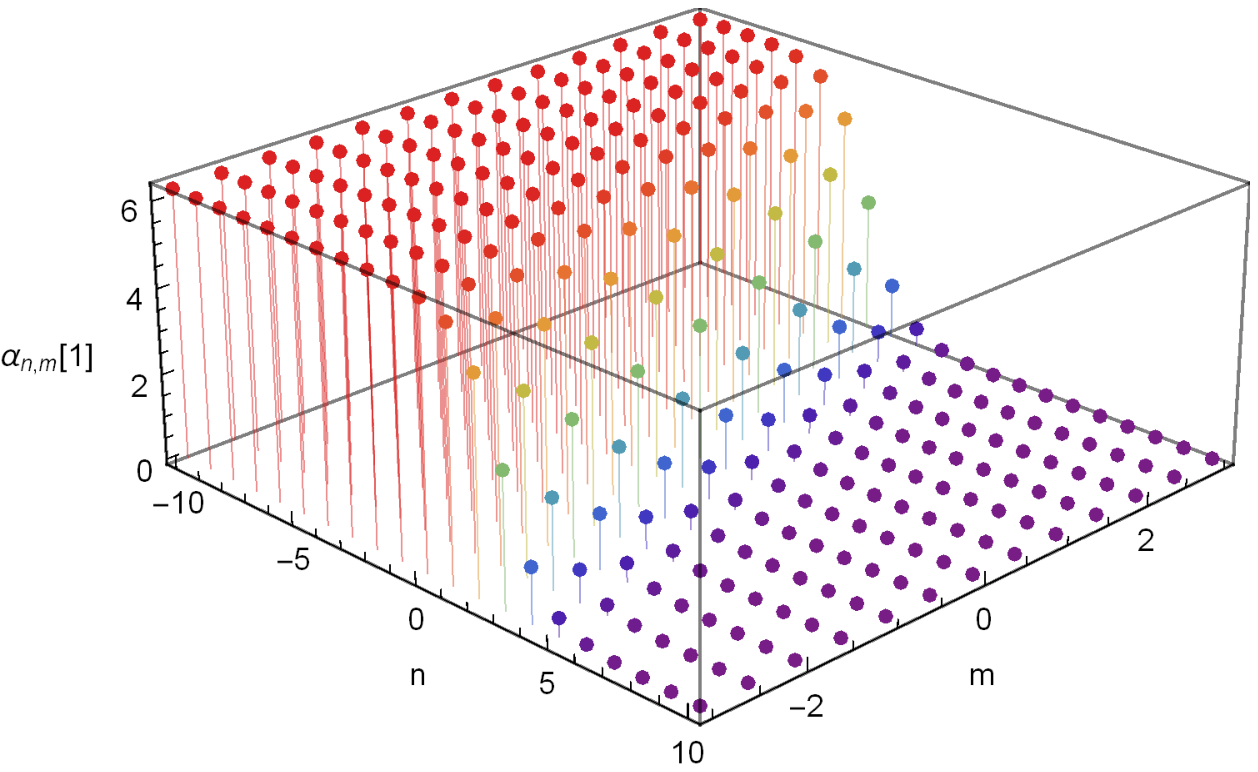}
       \caption{Propagation of one-kink soliton}
       \label{results1A}
        \end{subfigure}\quad
         \begin{subfigure}[b]{0.4\textwidth}
       \includegraphics[width=\textwidth]{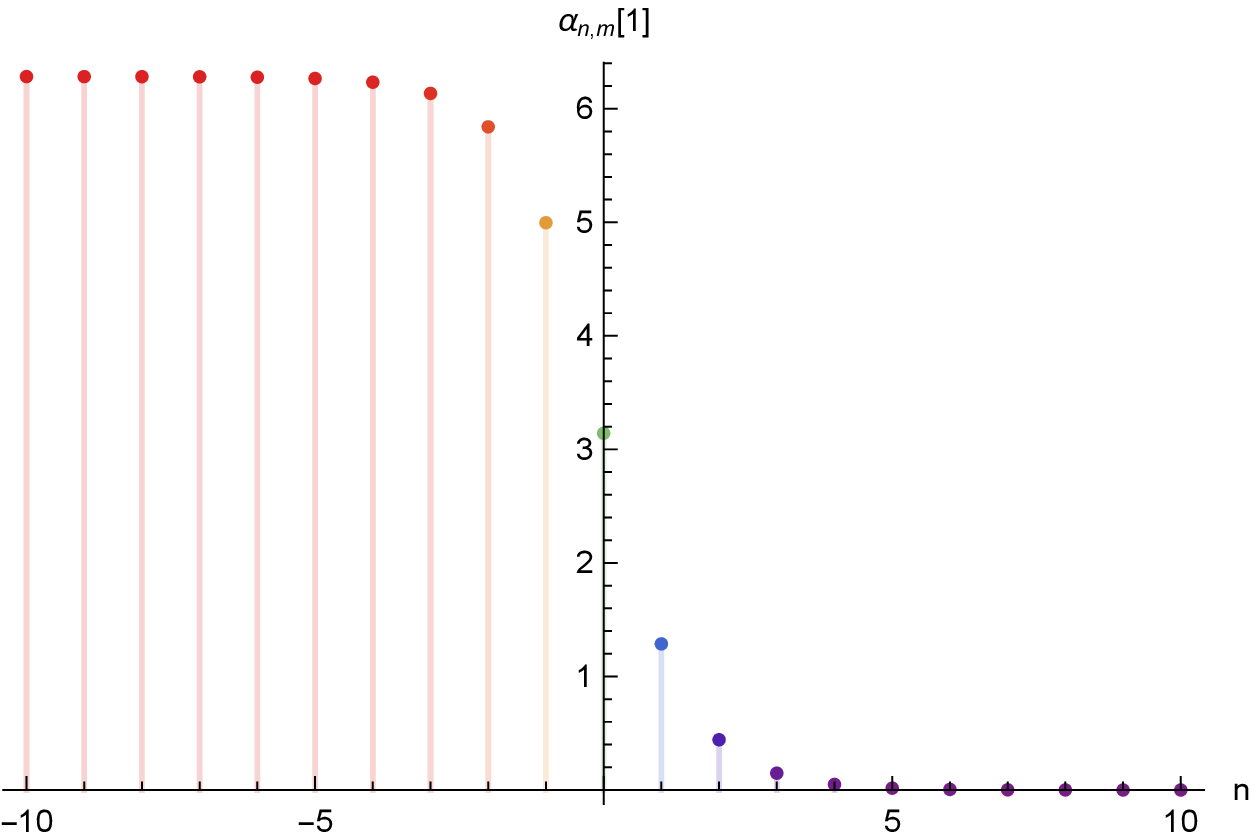}
        \caption{Snapshot of one-kink at $m=0$}
       \label{results1B}
    \end{subfigure}\\
         \caption{Propagation and snapshots of single-soliton solution (\ref{sg40}).}
    \label{onekink}
\end{figure}

In the zero background, that is $\alpha_{n,m}=0$, the linear system of difference-difference equations (\ref{sg6})-(%
\ref{sg7}) reduces to
\begin{eqnarray}
X_{n+1,m}&=&X_{n,m}+\lambda Y _{n,m},\quad\quad\quad Y_{n+1,m}=Y
_{n,m}+\lambda X
_{n,m}, \label{sg36}\\
X_{n,m+1}&=&X _{n,m}+\lambda ^{-1}\gamma Y _{n,m},\,\,\quad
Y_{n,m+1}=Y _{n,m}+\lambda ^{-1}\gamma X _{n,m}. \label{sg37}
\end{eqnarray}%
The solution to the linear system of difference-difference equations (\ref{sg36})-(\ref%
{sg37}) is given by%
\begin{eqnarray}
X _{n,m}&=&A\left( 1+\lambda\right) ^{n}\left( 1+\frac{\gamma }{%
\lambda}\right)^{m}  +B\left( 1-\lambda\right) ^{n}\left( 1-\frac{\gamma }{%
\lambda}\right)^{m} ,  \label{sg38} \\
Y_{n,m} &=&A\left( 1+\lambda \right) ^{n}\left( 1+\frac{\gamma }{%
\lambda}\right)^{m}  -B\left( 1-\lambda \right) ^{n}\left( 1-\frac{\gamma }{%
\lambda}\right)^{m}  ,  \label{sg39}
\end{eqnarray}%
where $A$ and $B$ are constants also known as plane wave factors.
Here
\begin{eqnarray}
X _{n,m} ^{(k)}&=&A_{k}\left( 1+\lambda_{k} \right) ^{n}\left( 1+\frac{\gamma }{%
\lambda_{k}}\right)^{m}+B_{k}\left( 1-\lambda_{k} \right) ^{n}\left( 1-\frac{\gamma }{%
\lambda_{k}}\right)^{m} ,  \label{sg38One} \\
Y _{n,m}^{(k)} &=&A_{k}\left( 1+\lambda_{k} \right) ^{n}\left( 1+\frac{\gamma }{%
\lambda_{k}}\right)^{m} -B_{k}\left( 1-\lambda_{k} \right) ^{n}\left( 1-\frac{\gamma }{%
\lambda_{k}}\right)^{m} ,  \label{sg39One}
\end{eqnarray}%
denotes  particular solution set to difference-difference equations (\ref{sg36})-(\ref%
{sg37}) at $\lambda=\lambda_{k}$. In order to obtain an explicit
expression of one-soliton substitute $X _{n,m} ^{(1)}$ and $Y
_{n,m}^{(1)}$
from (\ref{sg38One})-(\ref{sg39One}) in expression (\ref{sg16}), we obtain%
\begin{eqnarray}
\alpha_{n,m}\left[ 1\right]&=&2\iota \ln \frac{A_{1}\left(
1+\lambda_{1}
\right) ^{n}\left( 1+\frac{\gamma }{%
\lambda_{1}}\right)^{m}-B_{1}\left( 1-\lambda_{1} \right) ^{n}\left( 1-\frac{\gamma }{%
\lambda_{1}}\right)^{m} }{A_{1}\left( 1+\lambda_{1} \right)
^{n}\left( 1+\frac{\gamma }{%
\lambda_{1}}\right)^{m}+B_{1}\left( 1-\lambda_{1} \right) ^{n}\left( 1-\frac{\gamma }{%
\lambda_{1}}\right)^{m} }.\label{sg40}
\end{eqnarray}%
For the choice $A_1=1, B_1=\iota$, we obtain
\begin{eqnarray}
\alpha_{n,m}\left[ 1\right]&=&2\iota\ln \frac{X _{n,m} ^{(1)*} }{X _{n,m} ^{(1)}}=4\tan^{-1}\left[\left(\frac{1+\lambda_{1}}{1-\lambda_{1}}
 \right)^{n}\left(\frac{\lambda_{1}+\gamma}{\lambda_{1}-\gamma} \right)^{m}\right],\label{sg40A}
\end{eqnarray}%
which is analogous to the expression of one-kink soliton solution
obtained in \cite{dsr22} but more general one as it follows
continuum limits. The dynamics of one-kink solution is displayed in
Figure (\ref{onekink}) for $\lambda_1=0.2$ and $\gamma=0.25$. It is
clear from Figure (\ref{onekink}), that one-kink soliton
(\ref{sg40A}) has wave span of $2\pi$ and inflexion point trace is
given by $
n(m)=\frac{\tanh^{-1}\left(\frac{\gamma}{\lambda_{1}}\right)}{\tanh^{-1}\left(\lambda_{1}\right)}$,
whereas the value of $\alpha_{n,m}[1]$ and slope at inflexion point
are  $\pi$ and
$4\arctan\left(\frac{1+\lambda_{1}}{1-\lambda_{1}}\right)$
respectively (For more details on kink dynamics see e.g.
\cite{dsr22}-\cite{dsr23}). The one-kink soliton solution
(\ref{sg40A}) may be helpful in describing the nonlinear dynamics of
double helix DNA molecule as well as other important physical
systems for example propagation of optical soliton in  nematic
liquid crystals (NLCs).
Under continuum limit as defined by (\ref{continuum1}), expression
(\ref{sg40A}) yields
\begin{eqnarray}
\alpha_{n}(t)\left[ 1\right]&=&4\tan^{-1}\left[\left(\frac{1+\lambda_{1}}{1-\lambda_{1}} \right)^{n}\exp \left( \frac{2\gamma }{%
\lambda_{1}}t\right)\right],\label{contonesoliton}
\end{eqnarray}%
which coincides with expression of one-kink soliton solution for
sd-SG equation \cite{Yasir2019}.

Again under continuum limit (\ref{continuum3}) and
$\lambda_{1}\rightarrow \mathcal{X}\lambda_{1}$, above expression
(\ref{contonesoliton}) reduces to
\begin{eqnarray}
\alpha(x,t)\left[ 1\right]&=&4\tan^{-1}\left[\exp \left( 2\lambda_{1}x+\frac{2\gamma }{%
\lambda_{1}}t\right)\right],\label{contonesoliton}
\end{eqnarray}
which represents kink soliton solution for SG equation (\ref{sg1}).

Now, using (\ref{sg38One})-(\ref{sg39One}) in (\ref{Twosg16}) for
$A_{k}=1, B_{2k-1}=B_{2k}^{*}=\iota$, the explicit form of
double-soliton solution can be expressed as
\begin{eqnarray}
\alpha_{n,m}\left[ 2\right]&=&4
\tan^{-1}\left(\frac{\left(\frac{\lambda _1+\lambda _2}{\lambda
_1-\lambda _2}\right)
   \left(\frac{1+\lambda _1}{1-\lambda _1}\right)^{n}\left(\frac{\lambda
_1+\gamma}{\lambda _1-\gamma}\right)^{m}-\left(\frac{\lambda
_1+\lambda _2}{\lambda _1-\lambda _2}\right)\left(\frac{1+\lambda
_2}{1-\lambda _2}\right)^{n}\left(\frac{\lambda _2+\gamma}{\lambda
_2-\gamma}\right)^{m}}{1+
   \left(\frac{1+\lambda _1}{1-\lambda _1}\right)^{n}\left(\frac{\lambda
_1+\gamma}{\lambda _1-\gamma}\right)^{m}\left(\frac{1+\lambda
_2}{1-\lambda _2}\right)^{n}\left(\frac{\lambda _2+\gamma}{\lambda
_2-\gamma}\right)^{m}}\right).\label{TWOEXP}
\end{eqnarray}

\begin{figure}[H]
   \centering
    \begin{subfigure}[b]{0.45\textwidth}
       \includegraphics[width=\textwidth]{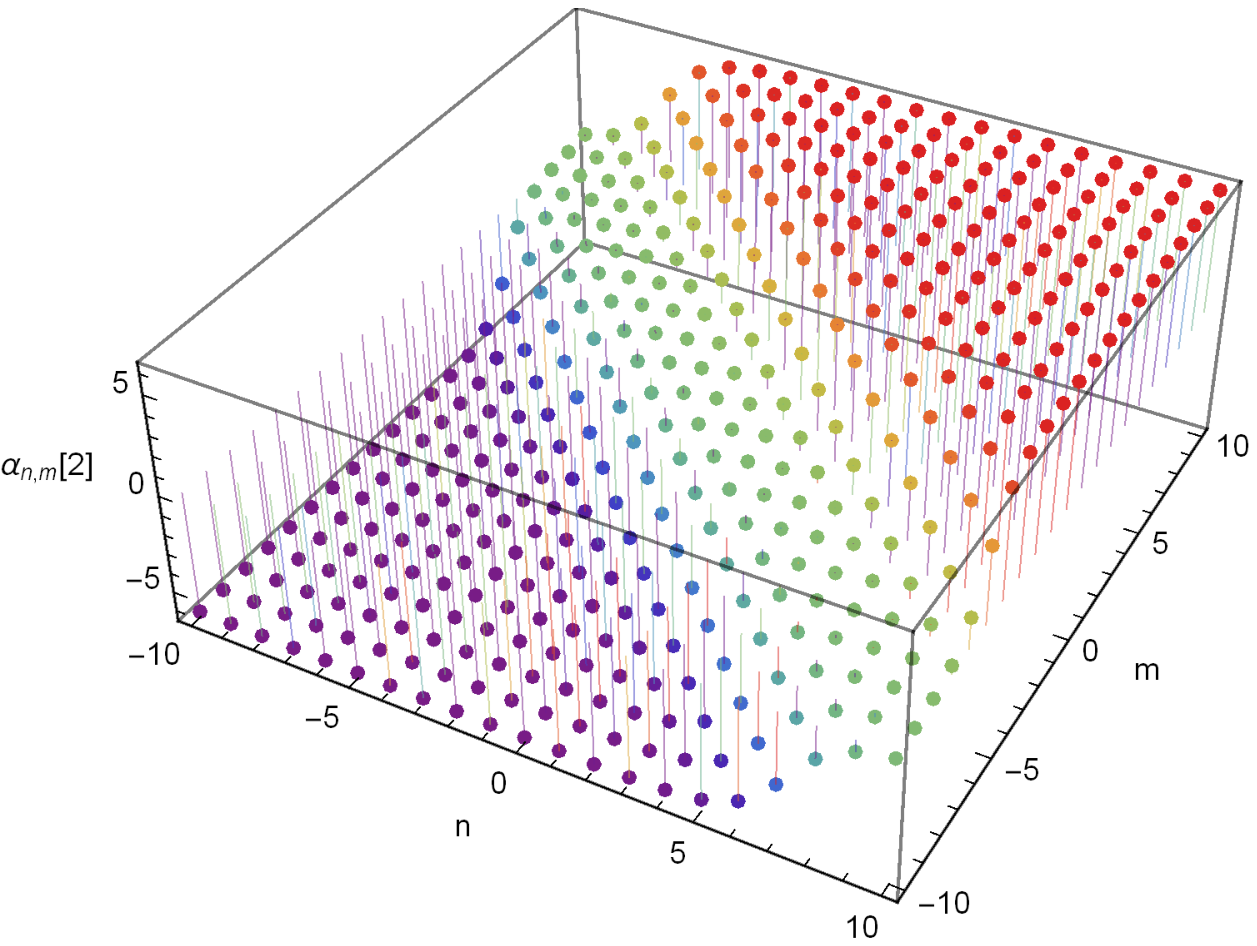}
       \caption{Propagation of double-kink}
       \label{results1A}
        \end{subfigure}\quad
         \begin{subfigure}[b]{0.4\textwidth}
       \includegraphics[width=\textwidth]{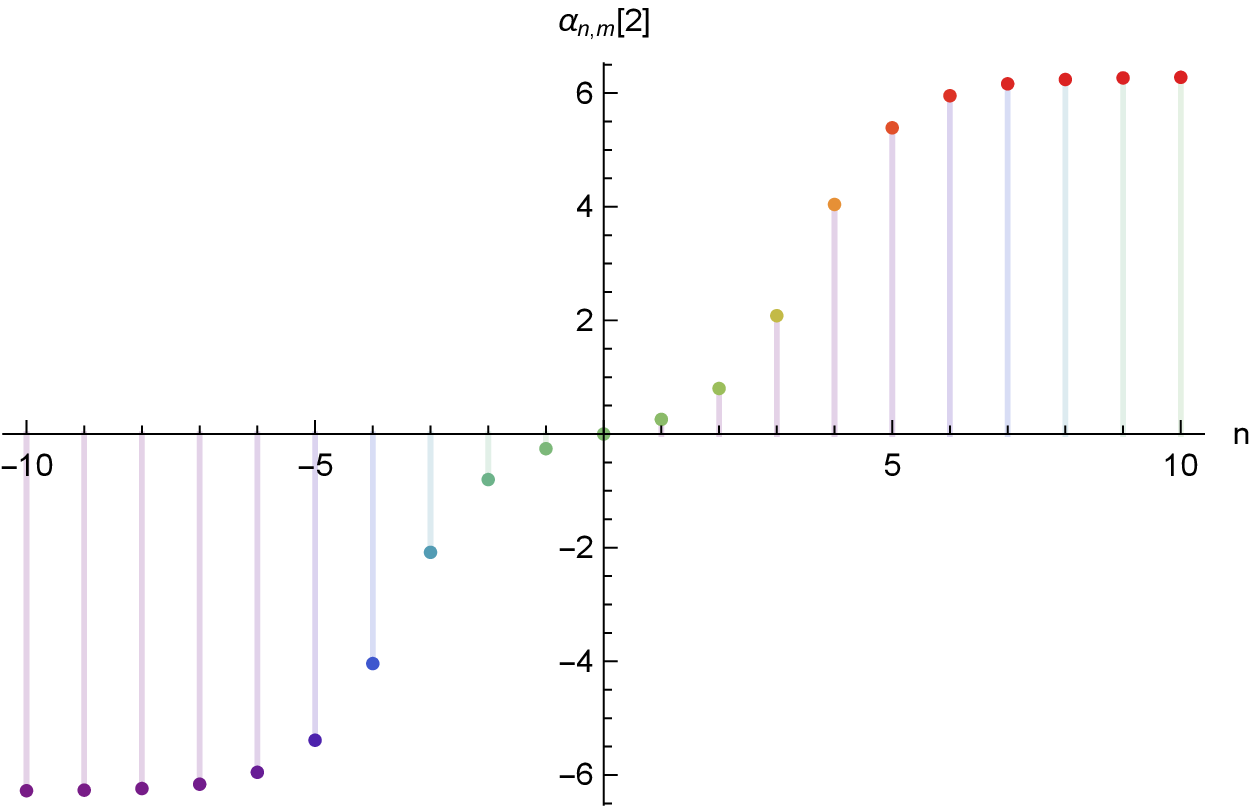}
        \caption{Snapshot of double-kink at $m=0$}
       \label{results1B}
    \end{subfigure}
     \caption{Propagation and snapshots of double-soliton solution (\ref{TWOEXP})}
    \label{twokinkkink}
\end{figure}

\begin{figure}[H]
   \centering
    \begin{subfigure}[b]{0.45\textwidth}
       \includegraphics[width=\textwidth]{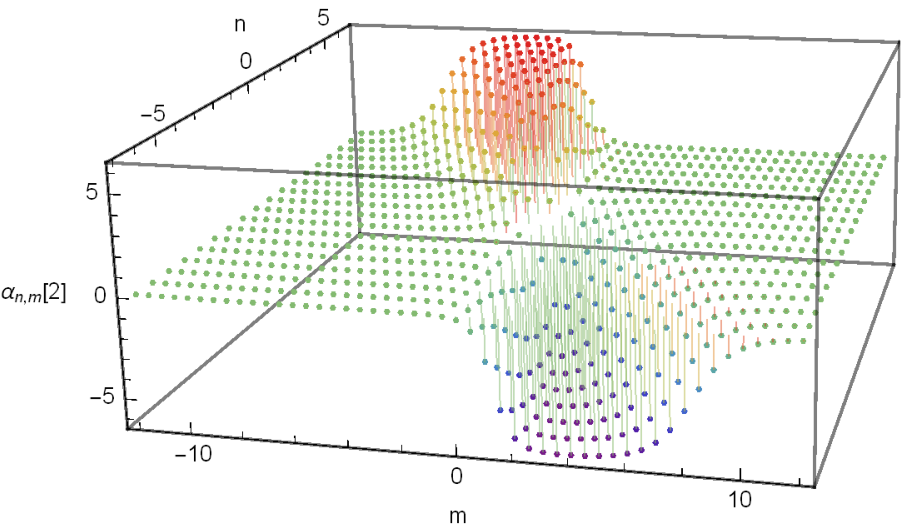}
       \caption{Propagation of kink-antikink}
       \label{results1AA}
        \end{subfigure}\quad
         \begin{subfigure}[b]{0.45\textwidth}
       \includegraphics[width=\textwidth]{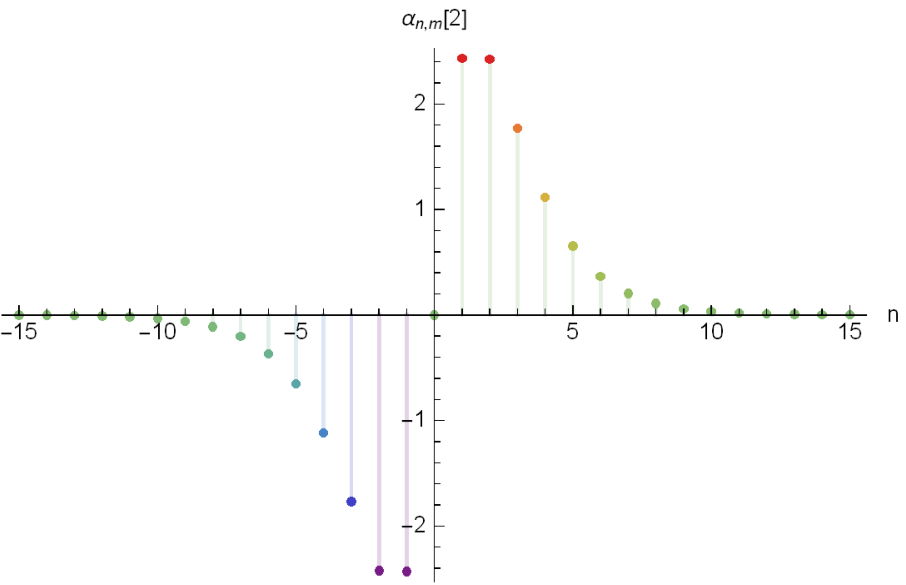}
        \caption{Snapshot of kink-antikink at $m=0$}
       \label{results1BB}
    \end{subfigure}
     \caption{Propagation of kink-anti kink interaction (\ref{TWOEXP})}
    \label{kinkantikink}
\end{figure}

Under continuum limits (\ref{continuum}), expression (\ref{TWOEXP})
yields the double-kink solution of the continuous SG equation
(\ref{sg1})
\begin{eqnarray}
\alpha(x,t)\left[ 2\right]&=&4\tan^{-1}\left[\left(\frac{\lambda
_1+\lambda _2}{\lambda _1-\lambda
_2}\right)\frac{\sinh\left(\frac{\left(\lambda_1-\lambda
_2\right)\left(\lambda _1 \lambda_2
   x-\gamma  t\right)}{\lambda _1
   \lambda _2}\right)
   }{\cosh\left(\frac{\left(\lambda _1+\lambda_2\right) \left(\lambda _1 \lambda _2
   x+\gamma
   t\right)}{\lambda_1
   \lambda _2}\right)}\right].\label{TWOEXPA}
\end{eqnarray}
The dynamics of kink-kink interaction is presented in Figure
(\ref{twokinkkink}) for $\lambda_{1}=-0.45,\,\lambda_{2}=0.5$ and
$\gamma=0.25$.  It is clear from Figure (\ref{twokinkkink}) when two
kinks approach each other with certain speed, they tend to repel one
another in the vicinity of $m=0$ and then two soliton bounce back
with velocities opposite to their initial velocities, as a result
kink-kink wave span goes from $-2\pi$ to $2\pi$. This is a case of
repulsive interaction.
An interaction of kink and anti-kink is depicted in Figure
(\ref{kinkantikink}) for $\gamma=0.25$,
$\lambda_{1}=0.3,\,\lambda_{2}=0.5$.
This Figure (\ref{kinkantikink}) shows when kink and antikink come
close to one another, they begin to accelerate in the neighborhood
of $m=0$ and then attempt to annihilate each other which corresponds
to the attractive case of interaction. After the interaction they
grow and separate as kink and anti-kink and move with same speed and
shape.

As $m\rightarrow\pm\infty$ two kinks are well separated and
expression (\ref{TWOEXP})  for $\lambda_{1}>\lambda_{2}$, can be
written as
\begin{eqnarray}
\alpha_{n,m}\left[ 2\right]&\cong&4
\tan^{-1}\left(\frac{\left(\frac{\lambda _1+\lambda _2}{\lambda
_1-\lambda _2}\right)
   \left(\frac{1+\lambda _1}{1-\lambda _1}\right)^{n}\left(\frac{\lambda
_1+\gamma}{\lambda _1-\gamma}\right)^{m}-\left(\frac{\lambda
_1-\lambda _2}{\lambda _1+\lambda _2}\right)\left(\frac{1+\lambda
_2}{1-\lambda _2}\right)^{n}\left(\frac{\lambda _2+\gamma}{\lambda
_2-\gamma}\right)^{m}}{1+
   \left(\frac{1+\lambda _1}{1-\lambda _1}\right)^{n}\left(\frac{\lambda
_1+\gamma}{\lambda _1-\gamma}\right)^{m}\left(\frac{1+\lambda
_2}{1-\lambda _2}\right)^{n}\left(\frac{\lambda _2+\gamma}{\lambda
_2-\gamma}\right)^{m}}\right),\label{TWOEXPASY}
\end{eqnarray}
the asymptotic response yields combination of kink and anti-kink
with phase shift that is due to their interaction at $m=0$
\begin{eqnarray}
\alpha_{n,m}\left[
2\right]&\cong&4\tan^{-1}\left[\left(\frac{\lambda _1+\lambda
_2}{\lambda _1-\lambda
_2}\right)\left(\frac{1+\lambda_{1}}{1-\lambda_{1}}
 \right)^{n}\left(\frac{\lambda_{1}+\gamma}{\lambda_{1}-\gamma} \right)^{m}\right]-\notag\\&&4\tan^{-1}\left[\left(\frac{\lambda _1-\lambda _2}{\lambda
_1+\lambda _2}\right)\left(\frac{1+\lambda_{2}}{1-\lambda_{2}}
 \right)^{n}\left(\frac{\lambda_{2}+\gamma}{\lambda_{2}-\gamma} \right)^{m}\right].\label{TWOEXPASYB}
\end{eqnarray}

\begin{figure}[H]
   \centering
    \begin{subfigure}[b]{0.5\textwidth}
       \includegraphics[width=\textwidth]{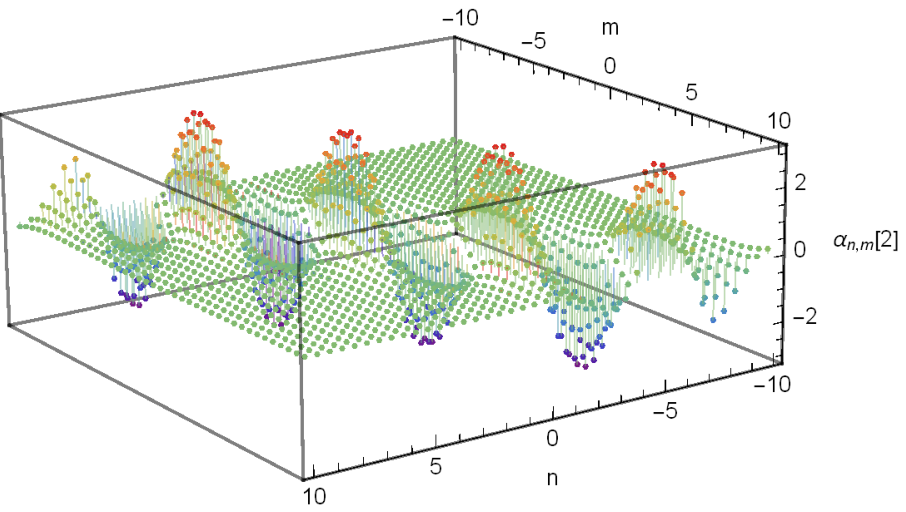}
       \caption{Propagation of single-breather solution}
       \label{onebreathA}
        \end{subfigure}\quad
         \begin{subfigure}[b]{0.4\textwidth}
       \includegraphics[width=\textwidth]{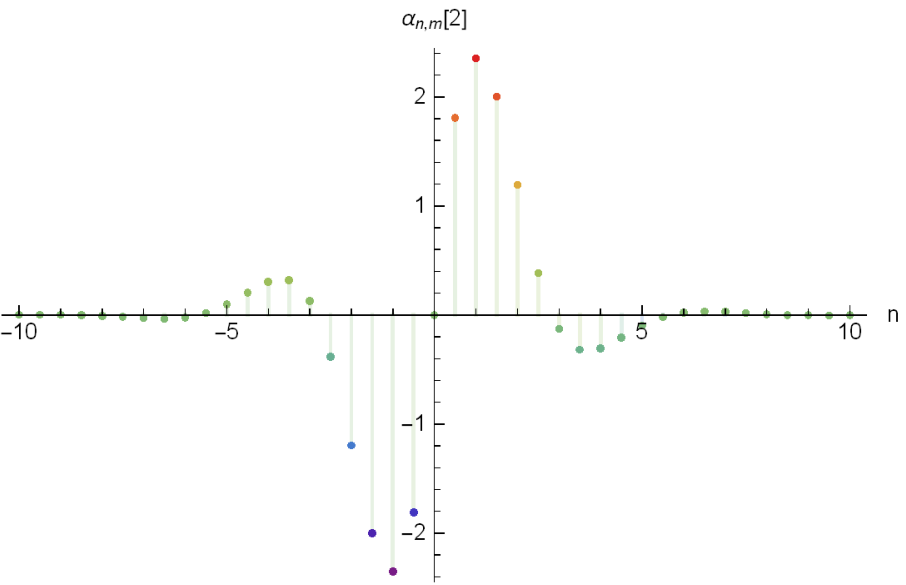}
        \caption{Snapshot of single-breather solution at $m=0$}
       \label{onebreathB}
    \end{subfigure}
     \caption{Propagation of single-breather solution}
    \label{onebreath}
\end{figure}

It is important to mention that the dSG equation (\ref{sg5}) also
admit another type of soliton solution known as doublet or breather,
that in general bound solution of soliton-antisoliton pair. We
obtain one-breather solution if we take
$\lambda_{2}=\lambda^{*}_{1}$ in (\ref{TWOEXP})
\begin{eqnarray}
\alpha_{n,m}\left[ 2\right]&=&4\imath
\tanh^{-1}\left(\frac{\Re(\lambda _1)}{\Im(\lambda
_1)}\left(\frac{1+\lambda _1}{1-\lambda
_1}\right)^{n}\left(\frac{\lambda _1+\gamma}{\lambda
_1-\gamma}\right)^{m}\frac{
   1-\mu_{n,m}}{1+\nu_{n,m}}\right),\label{BRETWOEXP}
\end{eqnarray}
with
\begin{eqnarray}
\mu_{n,m}&=&\left(\frac{1-2\iota\Im(\lambda_1)-|\lambda_1|^2}{1+2\iota\Im(\lambda_1)-|\lambda_1|^2}\right)^{n}\left(\frac{\gamma^2-2\iota\gamma\Im(\lambda_1)-|\lambda_1|^2}{\gamma^2+2\iota\gamma\Im(\lambda_1)-|\lambda_1|^2}\right)^{m},\notag\\
\nu_{n,m}&=&\left(\frac{1+2\Re(\lambda_1)+|\lambda_1|^2}{1-2\Re(\lambda_1)+|\lambda_1|^2}\right)^{n}\left(\frac{\gamma^2+2\gamma\Re(\lambda_1)+|\lambda_1|^2}{\gamma^2-2\gamma\Re(\lambda_1)+|\lambda_1|^2}\right)^{m}.\notag
\end{eqnarray}
Evolution of one-breather solution (\ref{BRETWOEXP}) for
$\lambda_{1}=0.5+0.5\iota$, is shown in Figure (\ref{onebreath}).

\begin{figure}[H]
   \centering
    \begin{subfigure}[b]{0.45\textwidth}
       \includegraphics[width=\textwidth]{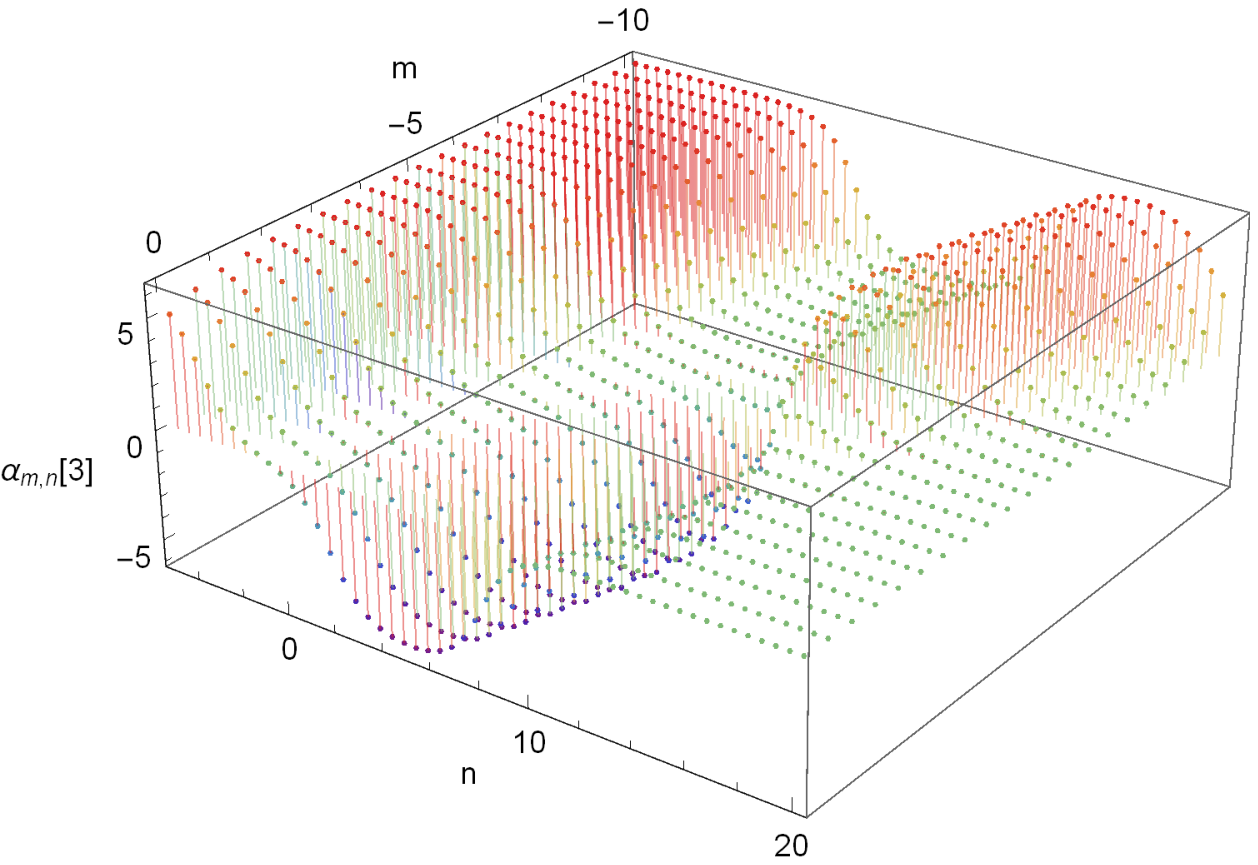}
       \caption{Propagation of interaction of two-kink and one-anti kink soliton}
       \label{results1A}
        \end{subfigure}\quad
         \begin{subfigure}[b]{0.4\textwidth}
       \includegraphics[width=\textwidth]{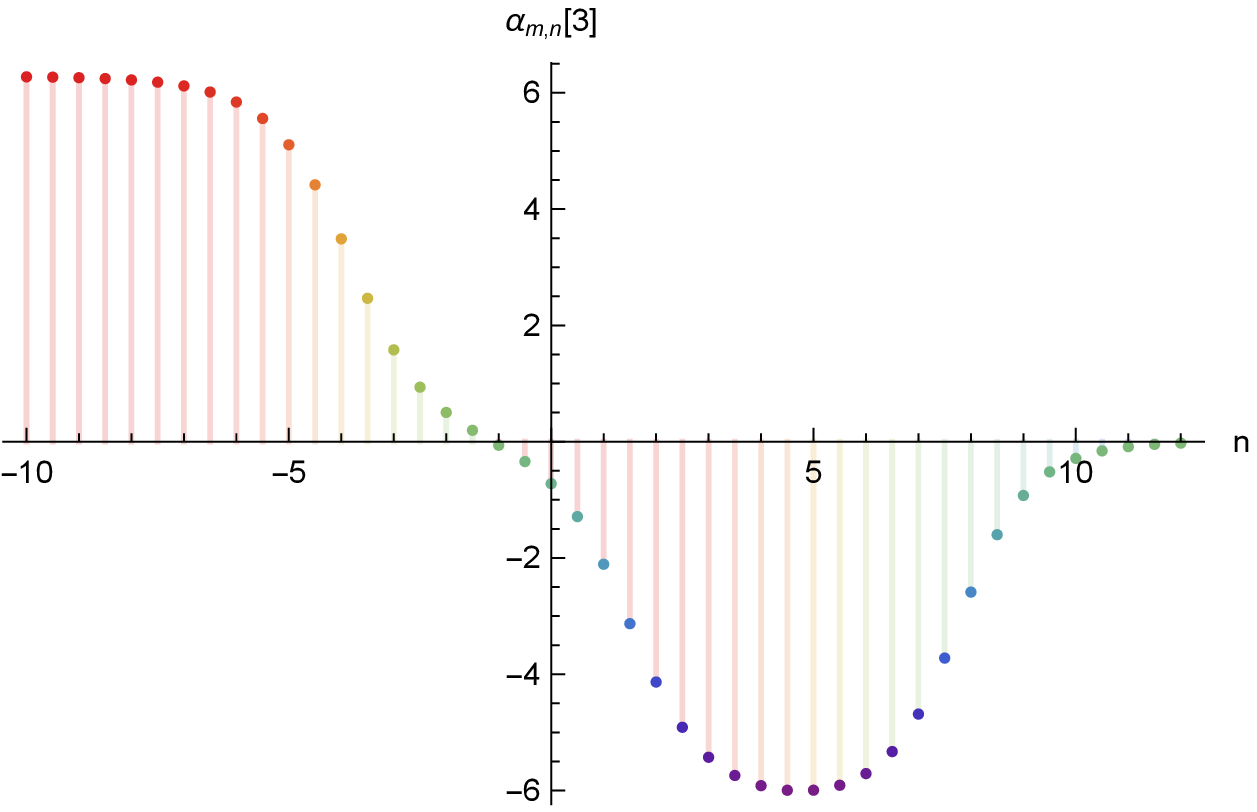}
        \caption{Snapshot of interaction of two-kink and one-anti kink soliton $m=0$}
       \label{results1B}
    \end{subfigure}
     \caption{Propagation and snapshot of triple-soliton solution (\ref{Threesg16}).}
    \label{Three1}
\end{figure}

Similarly, the triple-soliton solution can be obtained from
(\ref{Threesg16}). Figure (\ref{Three1}) represents interaction of
two-kink and anti-kink solutions for $\gamma=0.25$,
$A_{1}=A_{2}=A_{3}=0.1$, $2B_{1}=B_{2}/2=2B_{3}=\iota$ and
$\lambda_{1}=0.45, \lambda_{2}=0.5, \lambda_{3}=0.55$.
Finally, the quad-soliton solution is given as
\begin{eqnarray}
\alpha_{n,m}\left[4\right] &=&2\iota\ln\frac{\det\left(
                \begin{array}{cccccccc}
 \lambda_{1}^{3} X^{(1)}_{n,m}& \lambda_{1}^{2} Y^{(1)}_{n,m}& \lambda_{1}X^{(1)}_{n,m}&Y^{(1)}_{n,m}\\
 \lambda_{2}^{3} X^{(2)}_{n,m}&\lambda_{2}^{2} Y^{(2)}_{n,m}& \lambda_{2}X^{(2)}_{n,m}&Y^{(2)}_{n,m}\\
 \lambda_{3}^{3} X^{(3)}_{n,m}&\lambda_{3}^{2} Y^{(3)}_{n,m}& \lambda_{3}X^{(3)}_{n,m}&Y^{(3)}_{n,m}\\
 \lambda_{4}^{3} X^{(4)}_{n,m}&\lambda_{4}^{2} Y^{(4)}_{n,m}& \lambda_{4}X^{(4)}_{n,m}&Y^{(4)}_{n,m}
                \end{array}
              \right)}{\det\left(
                \begin{array}{cccccccc}
 \lambda_{1}^{3} Y^{(1)}_{n,m}& \lambda_{1}^{2} X^{(1)}_{n,m}& \lambda_{1}Y^{(1)}_{n,m}&X^{(1)}_{n,m}\\
 \lambda_{2}^{3} Y^{(2)}_{n,m}&\lambda_{2}^{2} X^{(2)}_{n,m}& \lambda_{2}Y^{(2)}_{n,m}&X^{(2)}_{n,m}\\
 \lambda_{3}^{3} Y^{(3)}_{n,m}&\lambda_{3}^{2} X^{(3)}_{n,m}& \lambda_{3}Y^{(3)}_{n,m}&X^{(3)}_{n,m}\\
 \lambda_{4}^{3} Y^{(4)}_{n,m}&\lambda_{4}^{2} X^{(4)}_{n,m}& \lambda_{4}Y^{(4)}_{n,m}&X^{(4)}_{n,m}
                \end{array}
              \right)} \label{FourEX}
\end{eqnarray}%

\begin{figure}[H]
   \centering
    \begin{subfigure}[b]{0.6\textwidth}
       \includegraphics[width=\textwidth]{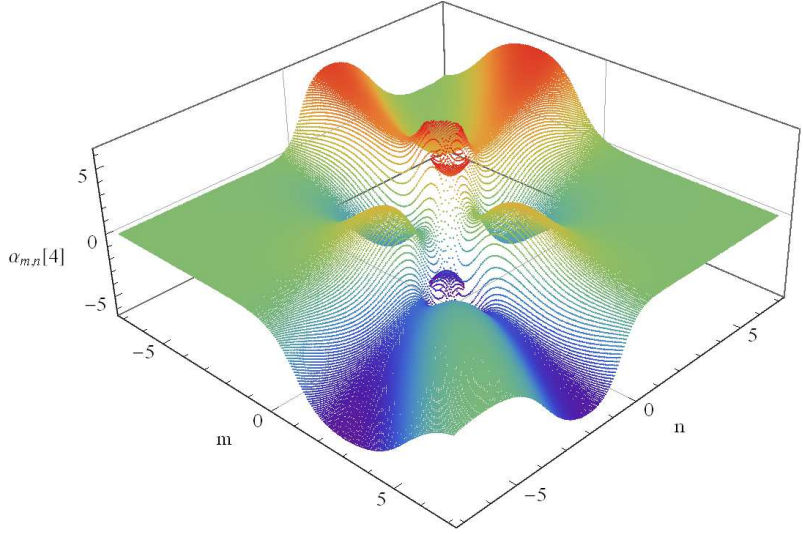}
       \caption{Interaction of double-kink and double-anti kink solutions}
       \label{FourA}
        \end{subfigure}\\
         \begin{subfigure}[b]{0.6\textwidth}
       \includegraphics[width=\textwidth]{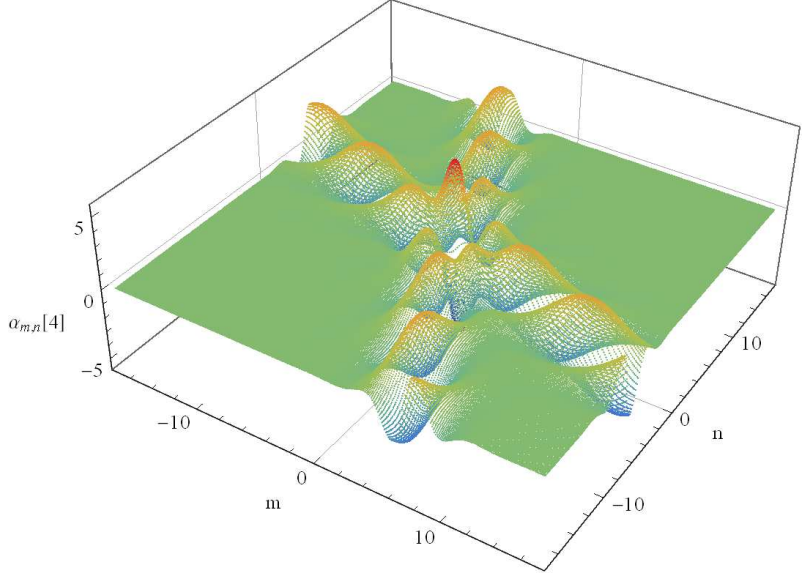}
        \caption{Interaction of double-breather solutions}
       \label{FourB}
    \end{subfigure}
     \caption{Dynamics of quad-soliton solution (\ref{FourEX})}
    \label{Four}
\end{figure}

Figure (\ref{Four}) represents dynamics of quad-soliton solutions
(\ref{FourEX}). Figure (\ref{FourA}) shows the interaction of
double-kink and double-anti kink solutions for $\gamma=0.25$,
$A_{k}=1, B_{2k-1}=B_{2k}^{*}=\iota$ and $\lambda_{1}=0.4,
\lambda_{2}=0.5, \lambda_{3}=0.6, \lambda_{4}=0.7$ and Figure
(\ref{FourB}) displays the interaction of double-breather solutions
for $\lambda_{1}=\lambda^{*}_{2}=0.7+0.7\iota,
\lambda_{3}=\lambda^{*}_{4}=0.2-0.2\iota.$

\section{Conclusions}
We have constructed multi-soliton solutions of dSG equation by
employing Darboux transformation and expressed our results as ratio
of ordinary determinants. We obtained explicit expressions of
single, double, triple and quad soliton solutions. We also obtained
single and double breather soliton solutions. Finally, we
illustrated different interactions of higher order soliton solutions
and breather solutions for dSG equation. Under continuum limit
results obtained in this paper reduce to multi-soliton solutions for
the classical sine-Gordon equation. It would be interesting to
explore dynamics of higher order degenerate solutions of discrete,
semi-discrete sine-Gordon equation. We shall study these solutions
in future.


\end{document}